# Visual Diagrammatic Queries in ViziQuer: Overview and Implementation

Jūlija OVČIŅŅIKOVA, Agris ŠOSTAKS, Kārlis ČERĀNS

Institute of Mathematics and Computer Science, University of Latvia
Raina blvd. 29, Riga, LV-1459, Latvia

julija.ovcinnikova@lumii.lv, agris.sostaks@lumii.lv, karlis.cerans@lumii.lv

**Abstract.** Knowledge graphs (KG) have become an important data organization paradigm. The available textual query languages for information retrieval from KGs, as SPARQL for RDF-structured data, do not provide means for involving non-technical experts in the data access process. Visual query formalisms, alongside form-based and natural language-based ones, offer means for easing user involvement in the data querying process. ViziQuer is a visual query notation and tool offering visual diagrammatic means for describing rich data queries, involving optional and negation constructs, as well as aggregation and subqueries. In this paper we review the visual ViziQuer notation from the end-user point of view and describe the conceptual and technical solutions (including abstract syntax model, followed by a generation model for textual queries) that allow mapping of the visual diagrammatic query notation into the textual SPARQL language, thus enabling the execution of rich visual queries over the actual knowledge graphs. The described solutions demonstrate the viability of the model-based approach in translating complex visual notation into a complex textual one; they serve as "semantics by implementation" description of the ViziQuer language and provide building blocks for further services in the ViziQuer tool context.

**Keywords:** Knowledge graphs, RDF, SPARQL, Visual queries, ViziQuer

## 1. Introduction

Knowledge graphs (Hogan et al., 2021), a data structuring paradigm, based on the graph data model of labelled nodes and edges, have become important for storing, aggregating, and linking information coming from various domains. Knowledge graphs are used on the web scale to support search engines. They are the way of information organization for Linked Open Data (WEB, d), where *DBPedia* (Auer et al., 2007) and *Wikidata* (Vrandečić and Krötzsch, 2014) are major data set examples. By describing the data on the level of semantic entities and relationships the knowledge graphs provide a higher and more user-friendly view of the data than, for instance, the relational databases (that describe the entities and their relations via primary and foreign keys).

Regarding knowledge graphs stored in the W3C standard RDF data format (Hayes and Patel-Schneider, 2014), the textual SPARQL language (WEB, b), also a W3C standard, is the common language for data querying. Besides the options for basic data pattern specification, SPARQL includes ways for rich query creation, involving, e.g., optional and negated fragments, aggregation, subqueries, and advanced property value conditions (way more than just equality and value intervals). The textual form of the SPARQL queries, however, complicates its use by various domain experts, and may not be the most convenient also for IT professionals.

There are a variety of tools and methods that can assist users in SPARQL query creation. There are tools such as YASGUI (Rietveld and Hoekstra, 2013) that provide help to the user in the creation of the textual form of the SPARQL query via syntax highlighting and text auto-completion facilities. Their strength is support for full SPARQL query functionality. Still, the range of users of the textual editors is limited to the SPARQL specialists, because to work with the textual SPARQL editor it is necessary to know both the SPARQL language and the data schema that is being queried.

The tools like *PepeSearch* (Vega-Gorgojo et al., 2016) and *WYSIWYQ* (Khalili and Merono-Penuela, 2017) offer means of SPARQL query construction via interaction with forms, where the values from the drop-down menus or radio buttons are chosen and text is entered in input fields. Their clear strength is ease to use for different user groups, however, they are much more limited with respect to the kinds of SPARQL queries that can be created via their support.

An interesting SPARQL query creation assistant is *SPARKLIS* (Ferré, 2017) that is based on the user interaction with the text snippets in controlled natural language, placing them together to obtain a textual description of the query. *SPARKLIS* can be praised for supporting most of the full SPARQL constructs, still, the used text snippets and the created controlled textual query formulations may seem to be somewhat artificial, and they may not always be the best way of describing and presenting the query structure.

Visual diagrammatic environments form another group of SPARQL query creation assistants, some of them (distantly) similar in style to the visual query builders for relational databases. Some of the visual diagram tools follow the UML-style (class-attribute-link) presentation of the data query backbone, such as *Optique VQs* (Soylu et al., 2018), *LinDA Query Designer* (WEB, c), *SPARQLGraph* (Schweigerr et al., 2014) and *ViziQuer* (cf. (Zviedris and Barzdins, 2011; Cerans et al., 2018b)). Others use a more detailed graphical presentation with attribute variables placed in separate graph nodes, such as e.g., *QueryVOWL* (Haag et al., 2015), *RDF Explorer* (Vargas et al., 2019), *GRUFF* (Aasman, 2017), and early works on *SEWASIE* (Catarci et al., 2003) and *NITELIGHT* (Russell and Smart, 2008).

While most of the visual diagrammatic query tools have such features as SPARQL endpoint querying, drag-and-drop, search, and auto-completion functionality, along with tool-specific features, the majority of the SPARQL query tools support only simple conjunctive queries and do not support the constructs such as sub-queries, aggregation, and advanced expressions. There is outer level aggregation possibility in *Optique VQs* and *LinDA Query Designer*, however.

The *ViziQuer* tool (Cerans et al., 2018b) with its initial notation presented in (Cerans et al., 2017) and (Cerans et al., 2018a) allows using the visual diagrammatic method for creating rich visual queries with optional and negated blocks, aggregation, subqueries, and advanced expressions, covering most of SPARQL 1.1 SELECT query constructs. This notation would allow a user that has started exploiting the visual query notation for simple query creation not to get stuck also when more complex tasks to be solved come at hand. There are also interesting options (cf. (Cerans et al., 2021a, 2022a, 2022b)) for generating a visual query from a given SPARQL query text (also with the rich query constructs supported) and usage examples of the notation (cf. (Cerans et al., 2019, 2021b)) reported.

The experience with the visual query creation in *ViziQuer* has shown, however, some necessary improvements and extensions of the notation, if compared to (Cerans et al., 2017) and (Cerans et al., 2018a), including, e.g., (i) explicit existence-checking edges, (ii) compartment-related filters, (iii) explicit grouping compartments, (iv) distinction among simple and non-duplicated condition implementation and (v) explicit support for *Wikidata* (Vrandečić and Krötzsch, 2014) language services.

This paper is the first one to provide an integrated account of the *ViziQuer* visual language with the extended features. The principal point of the paper, however, is to explain the principles and solutions behind the implementation of the *ViziQuer* visual query language, including the structures, techniques and algorithms used for the actual mapping of its complex visual structures into text-based ones of the SPARQL language.

The query translation uses the query abstract syntax tree (AST) that presents the query structure in accordance with the conceptual visual query components as nodes, links, data fields and conditions, as well as parsed textual expressions and resolved names of the data schema elements. When the AST is built from the technical query format natively supported by the diagramming engine, it is further on transformed into a scaffolding model for SPARQL query generation, from which the textual SPARQL query form is obtained.

The presented visual query implementation can be viewed as an experience story of applying a model-based conceptualization to a complex transformation task among radically different notations (a visual diagrammatic and a textual one) describing the same computational artefact (a query to retrieve the data from a knowledge graph).

Regarding the visual query notation itself, the provided query translation account can be seen as a "semantics by implementation" description that is provided here for the first time and includes details as visibility scope for introduced names and the order of blocks in the generated SPARQL query that, while being important for a wide range of queries, except the simplest ones, also has not been described before.

The *ViziQuer* software described in this paper is available online both as a playground environment[1] and an open-source project repository[2].

In what follows, Section 2 provides an overview of the *ViziQuer* visual query notation (what are queries that need to be translated into SPARQL). Section 3 outlines the query translation process and develops its basic structures of AST and the symbol table. Section 4 then describes the SPARQL query generation, including the SPARQL query generation model structure and the process of its construction, as well as the generation of the textual form of the SPARQL query. Section 5 concludes the paper.

## 2. ViziQuer Notation Overview

A visual query is created in the context of a data model that provides the vocabulary of entities, mapping an entity's local name and optional name prefix to the full entity URI, as well as stating the applicability, ordering and cardinalities of properties in the context of the model classes. In what follows, we shall demonstrate queries over a simple mini-hospital data schema, shown in Figure 1. The names of properties connecting the classes, if not specified, coincide with the target class name with a lowercase first letter. There is the default minimum and maximum cardinality 1 assumption for properties.

[1] https://viziquer.app

[2] https://github.com/LUMII-Syslab/viziquer

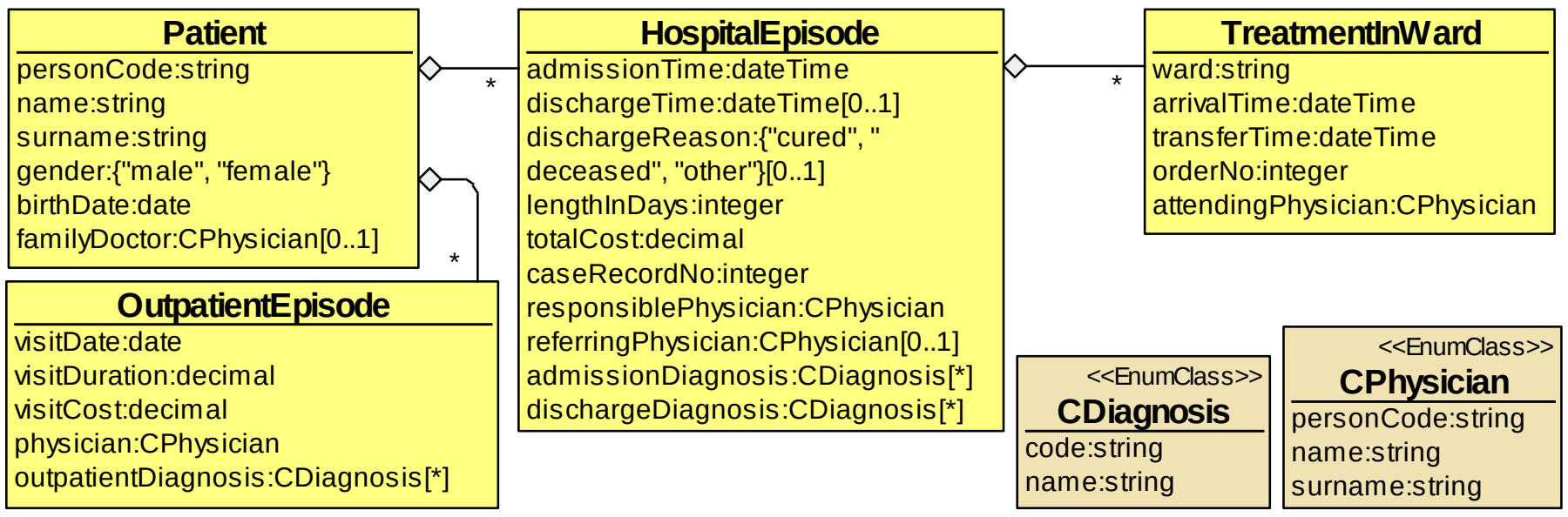

**Figure 1.** Example hospital schema (domain ontology) fragment (cf. (Barzdins et al., 2016; Cerans et al., 2018a))

## 2.1. Basic Visual Queries

A basic visual query (cf. (Zviedris and Barzdins, 2011; Cerans et al., 2017)) is a UML class diagram style graph with nodes describing data instances, edges describing their connections and fields forming the query selection list from the node instance model attributes and their expressions; every node can also specify the instance class and additional conditions on the instance. One of the graph nodes is the main query node (shown as an orange round rectangle in the concrete syntax); the structural edges (all edges except the reference ones, cf. Section 2.4) within the graph form its spanning tree with the main query node being its root.

Figure 2 shows an example basic visual query: *find 10 most expensive hospital episodes that last for at least 10 days, have a discharge reason specified and have a patient without any outpatient episode; list episode case record number, total cost and discharge reason, the patient's name and birth year, and the name of the referring physician if specified*.

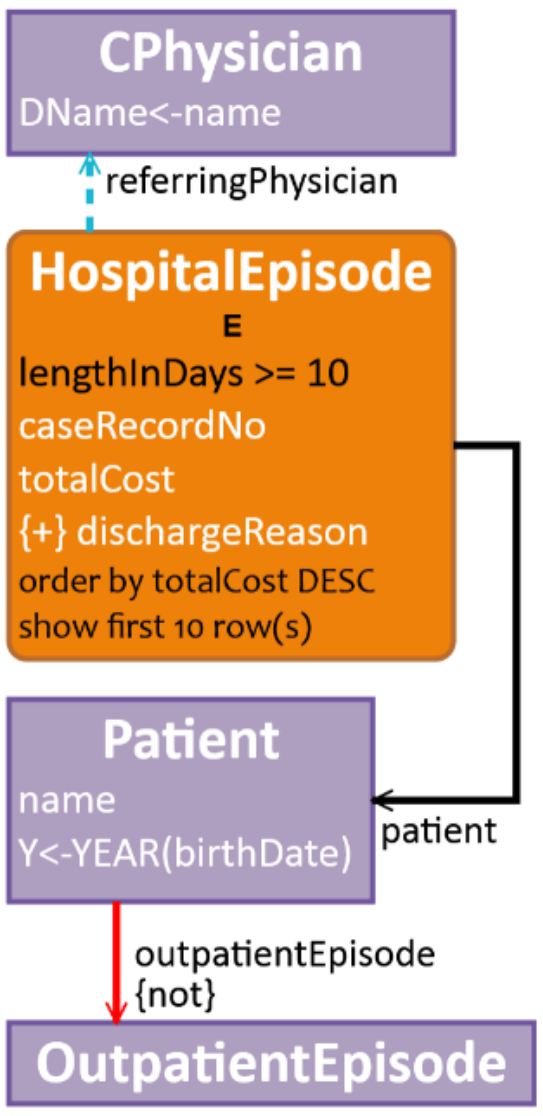

```
PREFIX : <http://lumii.lv/ontologies/2016/mini-bkus-en#>
PREFIX xsd: <http://www.w3.org/2001/XMLSchema#>
SELECT ?caseRecordNo ?totalCost ?dischargeReason
                              ?DName ?name ?Y WHERE{
  ?E a :HospitalEpisode.
  ?E :patient ?Patient.
  ?Patient a :Patient.
  OPTIONAL{?E :caseRecordNo ?caseRecordNo.}
  OPTIONAL{?E :totalCost ?totalCost.}
  ?E :dischargeReason ?dischargeReason.
  ?E :lengthInDays ?lengthInDays.
  OPTIONAL{?Patient :name ?name.}
  OPTIONAL{?Patient :birthDate ?birthDate.}
  OPTIONAL{ ?E :referringPhysician ?CPhysician.
    ?CPhysician a :CPhysician.
    OPTIONAL{?CPhysician :name ?DName.} }
  FILTER NOT EXISTS{
    ?Patient :outpatientEpisode ?OutpatientEpisode.
    ?OutpatientEpisode a :OutpatientEpisode. }
  BIND(YEAR(xsd:dateTime(?birthDate)) AS ?Y)
  FILTER(?lengthInDays >= 10)
} ORDER BY DESC(?totalCost) LIMIT 10
```

**Figure 2.** An example basic visual query and its translation into SPARQL

The links in the query have ascribed the property names (property paths are possible) from the data schema (property variables are allowed, as well). A link can be *required* (e.g., *patient* in Figure 2), *optional* (e.g., *referringPhysician*) and *negated* (e.g., *outpatientEpisode*). The query part behind an optional or a negation edge construction is seen as included in the optional or negation query block, respectively.

Each node in the query may (but is not required to) contain a class specification (e.g., *HospitalEpisode*, *CPhysician*, etc. in Figure 2).

For a node, it is possible to specify data fields, each holding a property or an expression to be selected in the query (together with an optional alias, e.g., *Y* or *DName* in Figure 2). By default, all selection fields in the query nodes are optional; a mark *{+}* marks the field to require values (as in *{+}dischargeReason*). There is also an option to mark an attribute as a helper (using mark *{h}*); that would mean finding the attribute but not including it in the selection list (such an attribute can still be referred to from other parts of the query).

Additionally, conditions over values of fields and other instance properties can be introduced (e.g., *lengthInDays >= 10*); they work as conditions over the selected data. There can also be data ordering expressions and slicing (limit, offset) specifications.

A query node can also have a data instance specification, either as a constant data resource (URI) or, more typically, as an instance name (e.g., *E* in Figure 2) that can be referred to from expressions in other parts of the query.

## 2.2. Aggregation and Grouping

The aggregated fields in the query can be placed in the main node of the query (and in the main nodes of "subqueries", introduced in Section 2.3) visually placed above the compartment for the class name (cf. *H_count* and *T_avg* in Figure 3). The body of such a field involves an aggregation function and its body expression – the aggregation subject that is typically the node instance itself (denoted by *(.)*, as in *count(.)*), or some its property value (as in *avg(totalCost)*), or even a more complex expression.

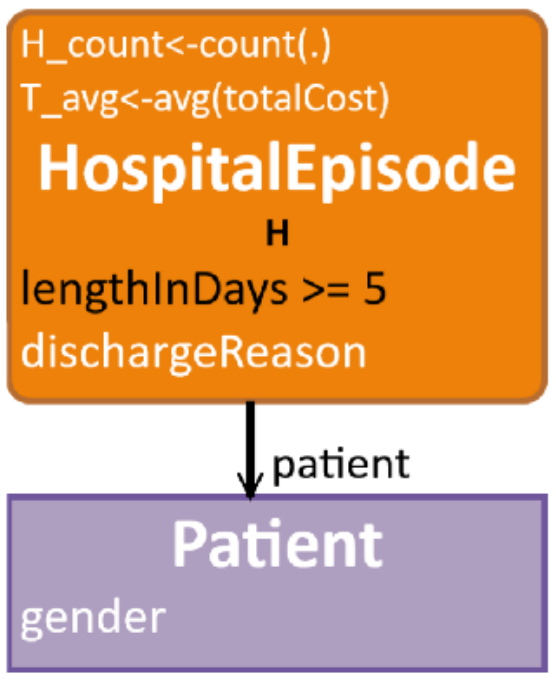

```
PREFIX : <http://lumii.lv/ontologies/2016/mini-bkus-en#>
SELECT ?dischargeReason ?gender (COUNT(?H) AS ?H_count)
                    (AVG(?totalCost) AS ?T_avg) WHERE{
  ?H a :HospitalEpisode.
  ?H :patient ?Patient.
  ?Patient a :Patient.
  OPTIONAL{?H :dischargeReason ?dischargeReason.}
  OPTIONAL{?H :totalCost ?totalCost.}
  ?H :lengthInDays ?lengthInDays.
  OPTIONAL{?Patient :gender ?gender.}
  FILTER(?lengthInDays >= 5)
} GROUP BY ?dischargeReason ?gender
```

**Figure 3.** An aggregate query example: *find count and average total cost for hospital episodes, grouped by their discharge reason and the patient's gender.*

If a query has non-aggregated attributes (selection fields), such as *dischargeReason* and *gender* in Figure 3, along with the aggregated ones, an implicit grouping over all non-aggregated selection fields is assumed (other grouping fields can be specified in explicit "group by" area, if necessary). Multiple aggregations are allowed within one node (as *H_count* and *T_avg* in Figure 3) if the aggregate expressions do not refer to "multi-valued" properties (the property *totalCost* in the hospital model has maximum cardinality 1), where that could distort the scope of other aggregations (if the "other aggregation" uses *min()* or *max()* function, the distortion does not happen, as it could for e.g., *count()* or *sum()*).

## 2.3. Subqueries

An important query construct in SPARQL 1.1, expanding substantially the query language capabilities, is that of a subquery, allowing to compute results that are further "injected" into the "outer"/"hosting" query. The visual query notation provides a concept of a subquery edge between a "host node" and a "linked node", allowing the fragment of the query that is behind the linked node to compute, within a subquery, some characteristics of the host node, that can be further referred to from the outer query. Technically, a subquery consists of the subquery edge (including its end element at the host node) and the entire query structure fragment behind the subquery edge.

A subquery would typically include an aggregation (possibly even as the sole explicit selection item). Figure 4 shows a simple query with subqueries.

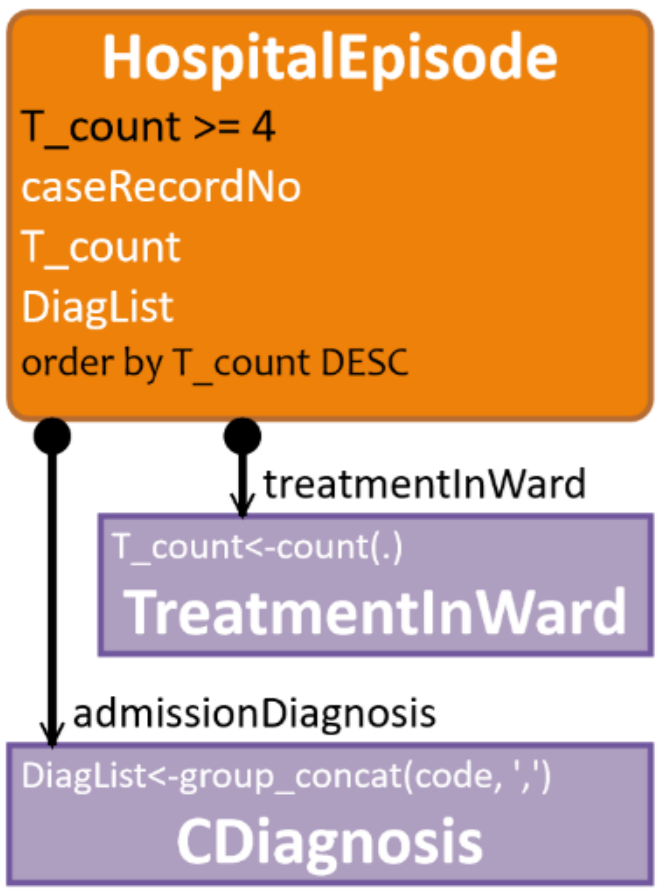

```
PREFIX : <http://lumii.lv/ontologies/2016/mini-bkus-en#>
SELECT ?caseRecordNo ?T_count ?DiagList WHERE{
  ?HospitalEpisode a :HospitalEpisode.
  {SELECT ?HospitalEpisode
        (COUNT(?TreatmentInWard) AS ?T_count) WHERE{
    ?HospitalEpisode :treatmentInWard ?TreatmentInWard.
    ?TreatmentInWard a :TreatmentInWard.}
    GROUP BY ?HospitalEpisode}
  {SELECT ?HospitalEpisode
        (GROUP_CONCAT(?code; SEPARATOR=',') AS ?DiagList) WHERE{
    ?HospitalEpisode :admissionDiagnosis ?CDiagnosis.
    ?CDiagnosis a :CDiagnosis.
    OPTIONAL{?CDiagnosis :code ?code.}}
    GROUP BY ?HospitalEpisode}
  OPTIONAL{?HospitalEpisode :caseRecordNo ?caseRecordNo.}
  FILTER(?T_count >= 4)
} ORDER BY DESC(?T_count)
```

**Figure 4.** Visual subquery example

In the simplest case, the subquery edge can be ascribed a property (as *treatmentInWard* and *admissionDiagnosis* in Figure 4) linking the host and linked nodes. The unconstrained (labelled by ++) and same-data (labelled by ==) edges (cf. Section 2.5) give further flexibility in the subquery construction.

A subquery link can be required, optional and negated (corresponding to SPARQL MINUS construct). A specific form of subquery is checking for the existence of a related resource, where the subquery would have no explicit selection fields, still the "existence checking" mode of the subquery needs to be asserted explicitly (to avoid the result set duplication in the case if the subquery pattern can be matched in multiple ways for a single host instance)[3]. Figure 5 provides an example. Note that the existence checking in simple cases can also be done by a textual condition, in the Figure 5 example situation such a condition would be *exists(hospitalEpisode.lengthInDays >= 5)* in the sole *Patient* node.

[3] This is different from the initial notation presentation in (Cerans et al., 2017)), where the subqueries had the "existence checking" semantics by default. The change does not affect the most common case of aggregated subqueries, where the duplication of the subquery result rows is not possible, however, the new option allows introducing also non-aggregated subqueries without implicit enforcing of the existence checking or distinct value selection mode.

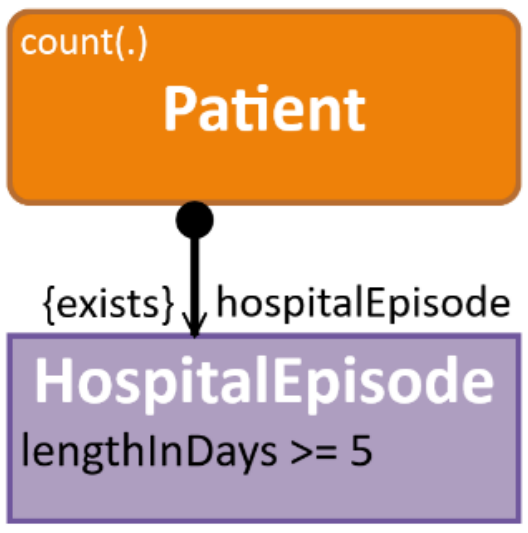

```
PREFIX : <http://lumii.lv/ontologies/2016/mini-bkus-en#>
SELECT (COUNT(?Patient) AS ?Patient_COUNT) WHERE{
  ?Patient a :Patient.
  FILTER(EXISTS{
    ?Patient :hospitalEpisode ?HospitalEpisode.
    ?HospitalEpisode a :HospitalEpisode.
    ?HospitalEpisode :lengthInDays ?lengthInDays.
    FILTER(?lengthInDays >= 5)
  })
}
```

**Figure 5.** Checking for the existence of a related resource in a subquery

The order by, limit and offset constructs are not allowed in the most typical subquery form of “local subqueries” meant for instance context property analysis[4]. There is a form of a “global subquery” in the visual notation, lifting these restrictions and making it clear that these queries cannot be used for context computation.

## 2.4. Reference Links

The visual notation allows creation of queries that have a richer than a tree-shaped structure. For this sake the reference[5] links are introduced that can add extra connections between the nodes of the query spanning tree defined by the query structure links, such as e.g., the join links and subquery links. A reference link can be required or negated, thus placing an extra assertion or its negation on the data to be selected (the optional reference links are not foreseen in the notation, as they would have logically void contents). Figure 6 shows an example of a negated reference link. In queries with nested block structure (due to e.g., subqueries or optional/negated structure links) a reference link can be allowed from a structurally deeper query block to a higher one.

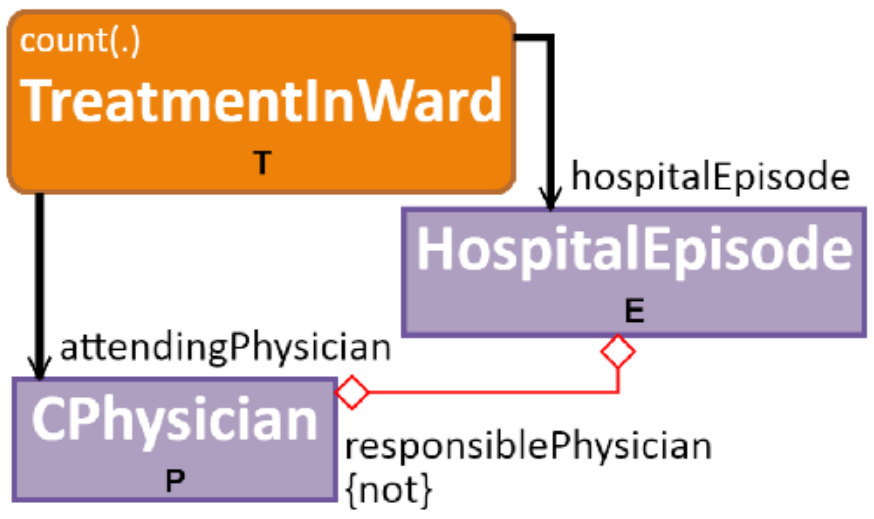

```
PREFIX : <http://lumii.lv/ontologies/2016/mini-bkus-en#>
SELECT (COUNT(?T) AS ?T_COUNT) WHERE{
  ?T a :TreatmentInWard.
  ?T :hospitalEpisode ?E.
  ?E a :HospitalEpisode.
  ?T :attendingPhysician ?P.
  ?P a :CPhysician.
  FILTER NOT EXISTS{?E :responsiblePhysician ?P.}
}
```

**Figure 6.** Example of a negated reference link: *count the treatments in ward that have the attending physician that is not the responsible physician for the wards’ hospital episode*[6].

---

[4] This is due to the subquery semantics in SPARQL, where the visual queries are translated into. It requires a subquery to be computed globally before embedding its results into the outer query.

[5] This type of link is called “condition” in (Cerans et al., 2017)) to reflect the intuition that these are added as triple conditions on top of an existing query structure.

[6] The definite articles are used in the query formulation since the maximum cardinality of the involved properties is 1.

## 2.5. Query Structure Extensions

The constructs introduced so far allow the creation of a wide range of visual queries, where the query structure matches the class and property structure of the data model. To expand the capabilities of the visual notation, the free edges (labelled by ++) and same-data edges (labelled by ==) are provided. A free edge just connects the nodes in the query structure without creating a data connection between them (typically, such a connection shall be created by reference links or cross-references to the node names in condition expressions). The same-instance edge, labelled by "==", provides a way of having the same data instance, represented by more than one node (this might be useful e.g., when more than one class name is to be specified for an instance, or when an instance node needs to be present both within a hosting query and a subquery). An alternative to using the "==" notation would be using the free "++" edge between the nodes while additionally using the same name of an instance in both examples. Figure 7 shows an example of using a negated free edge together with the subquery and reference link constructs (the *OutpatientEpisode O* needs to be brought into the subquery together with the *HospitalEpisode H* to have the name *H* visible at *O*).

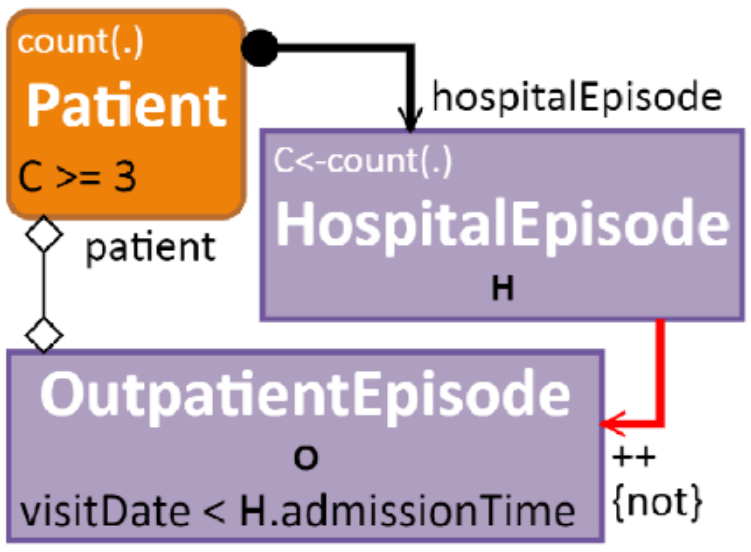

```
PREFIX : <http://lumii.lv/ontologies/2016/mini-bkus-en#>
SELECT (COUNT(?Patient) AS ?Patient_COUNT) WHERE{
  ?Patient a :Patient.
  {SELECT ?Patient (COUNT(?H) AS ?C) WHERE{
    ?Patient :hospitalEpisode ?H.
    ?H a :HospitalEpisode.
    FILTER NOT EXISTS{
      ?O a :OutpatientEpisode.
      ?O :patient ?Patient.
      ?H :admissionTime ?admissionTime_H.
      ?O :visitDate ?visitDate.
      FILTER(?visitDate < ?admissionTime_H) }}
    GROUP BY ?Patient }
  FILTER(?C >= 3) }
```

**Figure 7**. Example of a free link (++): *Count the patients that have at least 3 hospital episodes and do not have any earlier outpatient episodes.*

Besides the free and same-data edges, there are also control (non-data) nodes, the unit node [ ] and the union node [ + ] that can be used for further query structuring. These nodes do not themselves describe any data instance, however, they can specify the fields, conditions, aggregations and orderings with the expressions referring to the data computed at other nodes (typical for [ ]). If a control node is directly under a data node in the query structure tree (typically for [ + ]), the data node properties and links can be used in the context of the control node, as well. Figure 8 and Figure 9 show examples of the usage of the unit and union nodes, respectively.

An important way of extending the query structure is also possible by the names defined in one place of the query (e.g., node or field aliases) and referring to them from other parts of the query. If the same name is used as an explicit alias in different query parts, the name would in all places refer to the same resource.

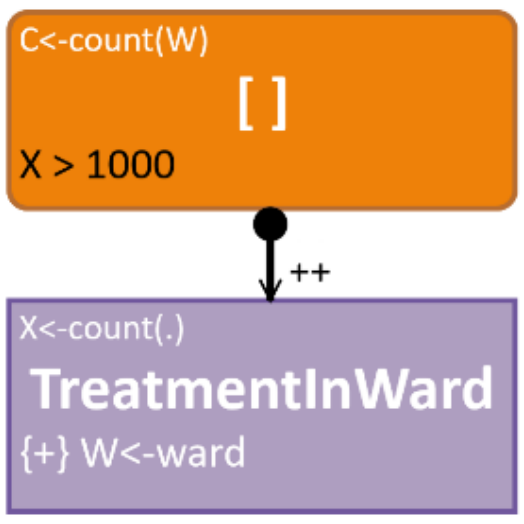

```
PREFIX : <http://lumii.lv/ontologies/2016/mini-bkus-en#>
SELECT (COUNT(?W) AS ?C) WHERE{
  {SELECT ?W (COUNT(?TreatmentInWard) AS ?X) WHERE{
    ?TreatmentInWard a :TreatmentInWard.
    ?TreatmentInWard :ward ?W.}
    GROUP BY ?W }
  FILTER(?X > 1000)
}
```

**Figure 8.** Unit node as a “wrapper” query: *Count all wards (attribute* ward *values) that have more than 1000 treatments in ward instances.* The wrapper queries can be used to emulate SPARQL HAVING construct, as well as to add further functionality (e.g., counting the matching values).

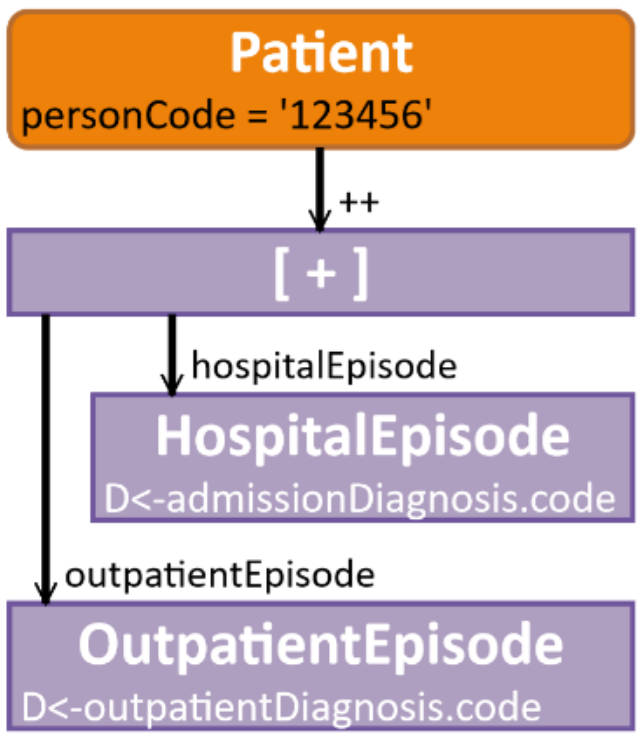

```
PREFIX : <http://lumii.lv/ontologies/2016/mini-bkus-en#>
SELECT ?D WHERE{
  ?Patient a :Patient.
  {?Patient :hospitalEpisode ?HospitalEpisode.
    ?HospitalEpisode a :HospitalEpisode.
    OPTIONAL{?HospitalEpisode :admissionDiagnosis/:code ?D.}
  }
  UNION {
    ?Patient :outpatientEpisode ?OutpatientEpisode.
    ?OutpatientEpisode a :OutpatientEpisode.
    OPTIONAL{?OutpatientEpisode :outpatientDiagnosis/:code ?D.}
  }
  ?Patient :personCode '123456' .
}
```

**Figure 9.** Union node example: *Find the codes of diagnoses that are made for a given patient either as hospital episode admission diagnoses or as outpatient episode diagnoses*.

## 2.6. Expression Notation

The visual query notation allows using textual expressions both in conditions (as e.g., *lengthInDays >= 10* in Figure 2) and in field value computation descriptions (e.g., *YEAR(birthDate)* in Figure 2). In the case of simple queries, the expressions are mostly a comparison of an attribute value with a constant or applying a function to an attribute value. Still, for the sake of the query notation completeness, more complex expressions are permitted, as well; a user can use those, as his/her query building skills permit.

Generally, all expression constructs that are available in SPARQL are supported also by the *ViziQuer* visual query notation, with an important modification that instead of SPARQL variables formed by the prefix ‘?’ the expressions in visual notation use explicit names defined in the query (as instance or field names/aliases) or the names of the properties (property paths allowed, as well). A property name in an expression at a node stands for a resource or a literal linked to the node instance (i.e., the “value” of the property). If a property name needs to be referenced as a resource in an expression, prefix its name with the inverse apostrophe (e.g., as in *`rdf:type*).

In addition to the basic “modified SPARQL” expressions there are several custom shorthands introduced, e.g.:

- notations ‘~’ and ‘~*’ as options for infix REGEX specification (use *a ~ b* for *REGEX(a,b)* and *a ~* b* for *REGEX(a,b,’i’)* (case insensitive matching),
- *abc[i]* for the initial part of *abc* of length *i* (as e.g., in condition *abc[1]=’A’*),

- *abc@en* for selecting the attribute *abc* only if its language tag is *en* (use *abc@(en,de)* for selecting attributes according to multiple language tags), and
- UML-style notation *'.'* for navigation expressions (SPARQL style property paths, built by *'/'*, are available, as well).

There is a further constraint on using arithmetic operators +, -, * and / in expressions: they need to be surrounded by spaces (to avoid conflicts with their use in forming URIs and property path expressions). The expression grammar is considered in Section 3.3.

## 2.7. Exploratory Queries

There are options to place explicit variables in the positions of a class and a property within the visual notation; in this case, the notation with an explicit '*?*' prefix is to be used. Some examples of exploratory queries are displayed in Figure 10.

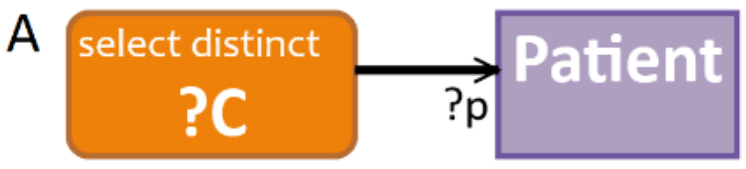

```
PREFIX : <http://lumii.lv/ontologies/2016/mini-bkus-en#>
SELECT DISTINCT ?C ?p WHERE{
  ?_C a ?C.
  ?_C ?p ?Patient.
  ?Patient a :Patient. }
```

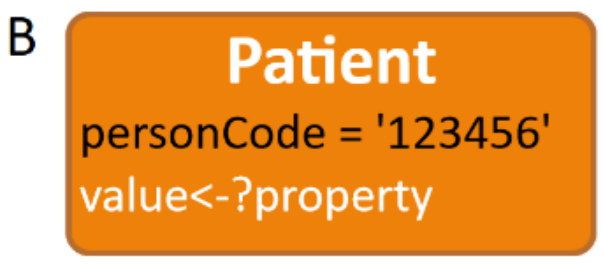

```
PREFIX : <http://lumii.lv/ontologies/2016/mini-bkus-en#>
SELECT ?property ?value WHERE{
  ?Patient a :Patient.
  OPTIONAL{?Patient ?property ?value.}
  ?Patient :personCode '123456' . }
```

**Figure 10.** Exploratory Queries: (a) *Find all class and property pairs* (C,p) *such that a link by* p *from some* C *instance goes to a patient*, and (b) *find all* (property,value) *pairs for a given person*.

If the class or property name to be found in the selection is not to be included in the query output, the corresponding variable name is to be preceded by '*??*' (can be used, e.g., for finding all resources connected to the given initial resource by any property, if the property name is not relevant).

The names for class and property variables can be referred to from other expressions within the query (to make a reference use the variable name without the '*?*' or '*??*' mark).

# 3. Visual Query Implementation

The implementation of the visual queries is provided within the frame of the *ViziQuer* visual query tool (cf. (Cerans et al., 2018b)) that offers means for visual query editing (***VQEditor***), translating the queries into SPARQL (***VQTranslator***) and executing the queries over given SPARQL endpoint (***VQExecutor***), as outlined in Figure 11.

The *ViziQuer* tool is implemented in the *ajoo* visual tool building platform (Sprogis, 2016) which allows the creation of visual domain-specific languages by offering means for rapid development of concrete syntax and visual editor for the language. Still, the implementation of the visual language semantics (translation of the visual queries into SPARQL in the case of the *ViziQuer* tool) lies outside the scope of the platform and is presented for the *ViziQuer* tool in this paper.

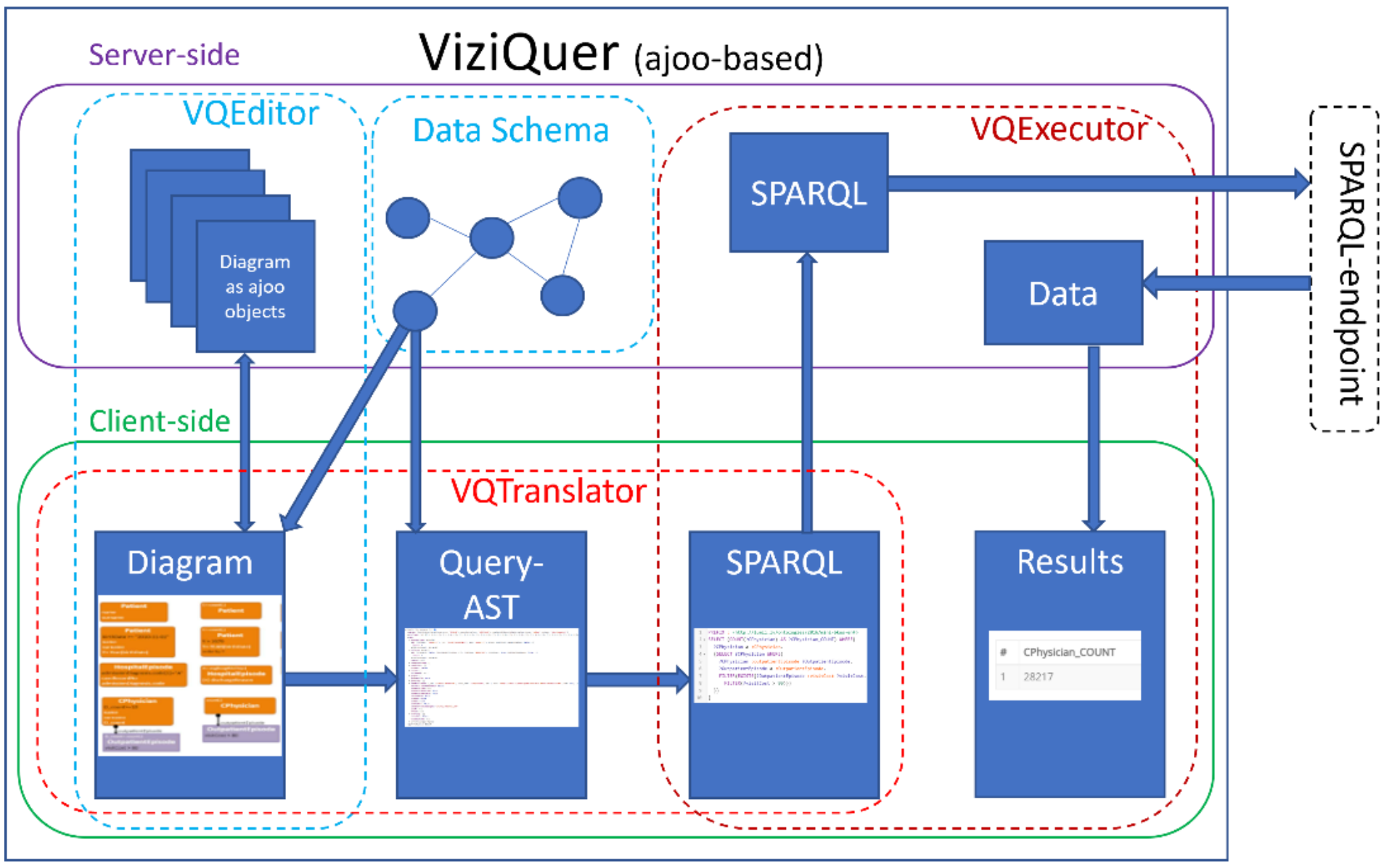

**Figure 11.** Architecture of *ViziQuer* and data flow of query editing and execution

The *ajoo* platform provides the representation of a visual diagram in the form of a graph of connected elements (boxes and lines), possibly with certain textual compartments (cf. *Presentation model* in Figure 12). There is also the *Type model* information available during the diagram runtime that allows discriminating among the roles that the different elements and compartments play in a diagram (cf. Figure 12) (cf. (Barzdins et al., 2007) for the explanation of the concept of runtime management of visual diagrams and their elements through their linked type/configuration elements).

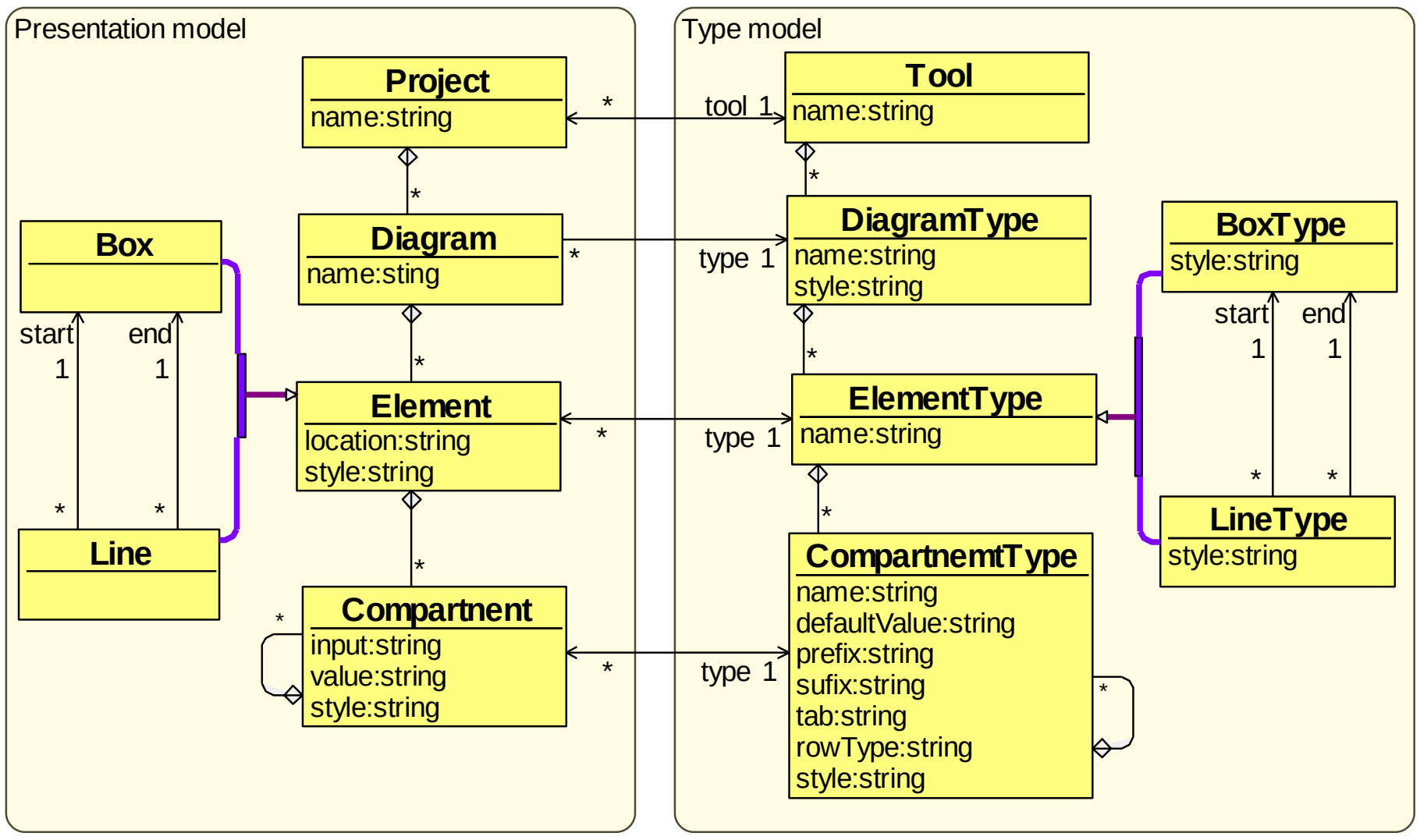

**Figure 12.** Metamodel of diagram presentation in *ajoo* tool

The visual query implementation starts with creating for it an abstract syntactic structure (Abstract Syntax Tree) that is based on semantic concepts of query composition as *Node*, (Attribute) *Field*, *Condition*, etc. (cf. Section 3.1) and that is used further (instead of the raw *ajoo*-structured representation) in the query generation: building the symbol table (Section 3.2), resolving the expression types (Sections 3.3 and 3.4), as well as generating the abstract internal model of the SPARQL query (Section 4), further on used to generate the textual SPARQL query form.

We note that the projects in the visual tool environment also have options for setting the parameters (described in Section 3.5) that can influence the query implementation.

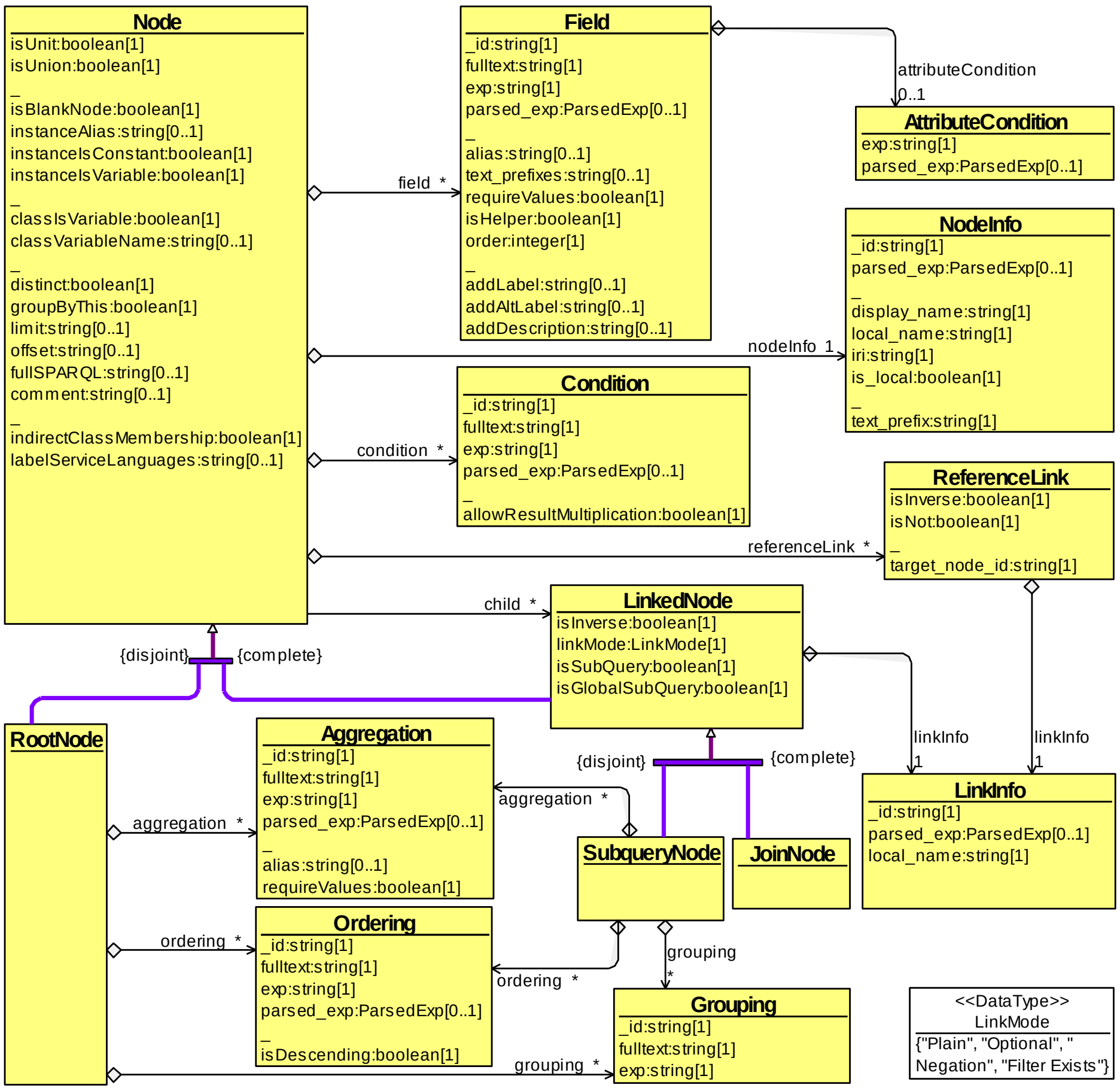

**Figure 13.** ViziQuer Graphical Query Abstract Syntax Tree Model

## 3.1. Abstract Syntax Tree Core Structure

The principal structure of the visual query implementation is that of the query Abstract Syntax Tree (AST) that encodes the query information in accordance with the conceptual visual query components as nodes, links, data fields and conditions, as well as parsed textual expressions and resolved names of the (background) data schema elements (the data schema lists the defined namespaces (one of them is the default namespace), as well as the data model entities – the classes and the properties, listing their full entity iri, as well as the namespace, local name, and display name; in the case of properties also their cardinality information can be made available (as it can influence the query generation)).

The AST structure is summarized in Figure 13. The central item of the AST structure is *Node*. Every node except the query root node (*RootNode*) belongs to the *LinkedNode* class, where both the node properties and the node attachment link to its parent node in the query structure spanning tree are specified. According to the incoming structure link type, the linked nodes are split into the *JoinNode* and *SubqueryNode* classes, where the subquery nodes can have further information items (e.g., aggregations, groupings, and orderings). The reference links (cf. Section 2.4) go into a separate class *ReferenceLink*, such a link is attached to its source node and describes the target node in textual form.

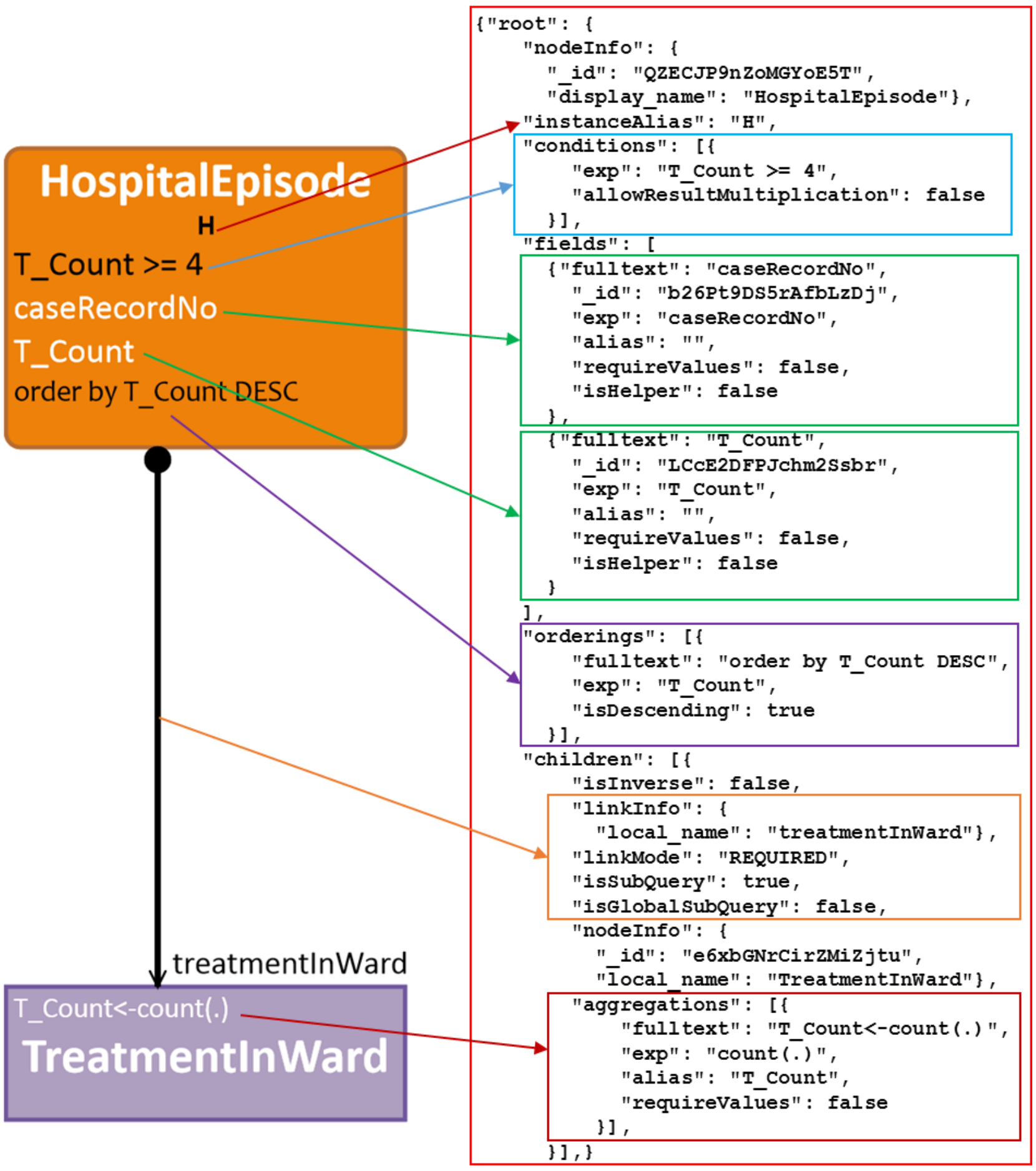

**Figure 14.** ViziQuer graphical query and its abstract syntax tree

Each graphical element (nodes, links, and node/link compartments: fields, conditions, aggregations, groupings, and ordering) has an *_id* attribute that technically identifies the element; for nodes, the *_id* attribute is stored in the related *NodeInfo* instance, and for links

– in the *LinkInfo* instance (so, for a *LinkedNode* instance there shall be both the node and its incoming structure link identifications available).

The references from the AST structure to the data model with the class and property information are limited to *local_name*, *iri*, *prefix* and *is_local* information in the *NodeInfo* class and *local_name* information in the *LinkInfo* class. There is, however, a *parsed_exp* attribute for all diagram elements that contains a parsed form of the textual expression placed in the element and the references to the data schema are provided from therein.

The AST is built from the diagram presentation model provided by the platform (cf. Figure 12) and the associated data schema information in two stages:

(i) Initial structure creation, builds all model information, except the *parsed_exp* attribute and the explicit references to the data model, and

(ii) Full AST creation based on the initial AST structure, involving expression parsing, name resolution (based on the data model information and on the names defined within the query) and creation of the symbol table (ST).

Figure 14 outlines an example of the ViziQuer graphical query in the initial AST notation, with expression parsing and schema element identification yet to be done.

## 3.2. The Symbol Table

The role of the symbol table is to identify names (identifiers) defined in a query, hold information on their typing and kind, as well as scope – accessibility in different nodes in the query (apart from the names described in the symbol table, the expressions at nodes can also use property and class names from the data model; these are introduced into the symbol table only if they make the entire expression field that can be further on referred to from other places in the query).

The symbol table is built initially based on the initial AST structure (not involving the schema information and parsed expressions) and is further enriched in parallel with the process of resolving the names appearing in the AST expressions (cf. Section 3.4).

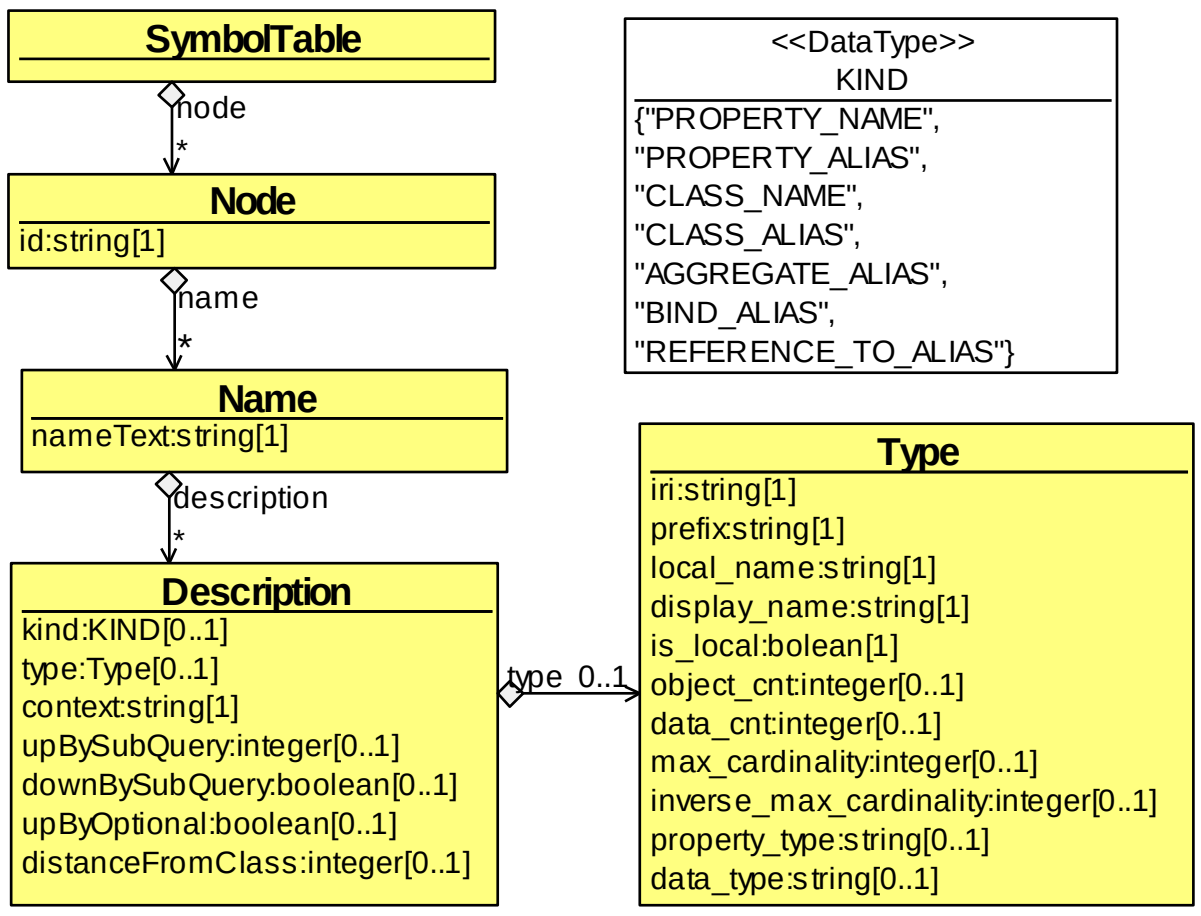

**Figure 15.** Structure of the Symbol Table

The structure of the symbol table is outlined in Figure 15. For each node the set of the names visible from its context is maintained, each name can have one or several descriptions denoting the possible different meanings of the name (the case with one description of a name is typical, however, several descriptions are possible, as well).

In the name description, the kind of the name shall have one of the following values:

- PROPERTY_NAME – the name is introduced in the query as a data model property.
- CLASS_NAME – the name is introduced in the query as a data model.
- PROPERTY_ALIAS – the name is defined in the query as an alias of a field holding a single property name or property path expression (can be implemented as a variable appearing in a SPARQL Basic Graph Pattern).
- BIND_ALIAS – the name is defined in the query as an alias of a field holding a more complex expression (requires BIND construct for implementation in SPARQL).
- CLASS_ALIAS – the name is defined in the query as a name of an instance described by a node.
- AGGREGATE_ALIAS – the name is defined in the query as an alias of an aggregated field (typically defined in some aggregated subquery).
- REFERENCE_TO_ALIAS – the name is defined in the query via a field expression that refers itself to an alias described elsewhere (such a reference can be used to extend the scoping of the alias also to places where the original alias definition would not be accessible (typically used with deeply nested subqueries)).

The accessibility of the names introduced in the query is summarized in Table 1. The contents of the symbol table reflect the availability of the names according to these rules.

**Table 1.** Summary of name scoping/visibility rules in visual queries

| | This node | Other nodes in the fragment | Down by optional, negation, union, or sub-union | Up by optional, union, or sub-union | Down by subquery | Up 1 level through a subquery link |
|---|---|---|---|---|---|---|
| Instance name (Class alias) | Yes | Yes | Yes | Yes (3) | Yes (4) | No (select as a field to export) |
| Property name (implicit alias) | Yes (1) | Yes (2) | No | Yes (3) | No | Yes (5) |
| Explicit alias (Property, Bind Reference to alias) | Yes (1) | Yes (2) | No | Yes (3) | No | Yes (5) |
| Aggregate alias | No | No | No | No | No | Yes (5) |

Legend:

(1) can be used in conditions and aggregations; usage in instance field expressions is after the name has been introduced.

(2) can be used in field expressions in nodes that are structurally above the node that introduces the name, as well as in conditions anywhere in the fragment.

(3) can be used as a value, not as a starting point of a navigation expression; (2) applies, as well.

(4) except in aggregate fields.

(5) the field must be selected from the subquery; after the move up through the subquery link to the host node, the field is visible within the current fragment only.

During the ***initial phase*** of construction (before the name resolution), the kind information for name descriptions shall have values CLASS_ALIAS and AGGREGATE_ALIAS (these are known from the initial AST structure) and shall use temporary markers UNRESOLVED_FIELD_ALIAS (in the case, if the field has an alias) and UNRESOLVED_NAME (all other cases).

Within the ***full symbol table*** (as constructed in Section 3.4) the name descriptions referring to the data model entities (i.e., being of the kind PROPERTY_NAME or CLASS_NAME) shall have *type* information describing the denoted entity (including e.g., the entity *iri* and *display_name*, as well as further data model information to be used in the user interface or the SPARQL query generation).

The *context* attribute of a description holds the AST *id* of the node where the name has been defined (where the definition of the alias appears; for names from the data model the current node *id* is stored here).

The remaining description attributes *upBySubquery*, *downBySubquery*, *upByOptional* and *distanceFromClass* assist in the symbol table building.

Figure 16 shows a fragment of the symbol table for the query of Figure 14.

```
"symbolTable": {
  "e6xbGNrCirZMiZjtu": {
    "H": [ {
        "kind": "CLASS_ALIAS",
        "type": {
          "iri": "http://lumii.lv/ontologies/2016/mini-bkus-en#HospitalEpisode",
          "prefix": "",
          "local_name": "HospitalEpisode",
          "display_name": "HospitalEpisode",
          "is_local": true },
        "context": "QZECJP9nZoMGYoE5T",
        "downBySubquery": true
      }]
  },
  "QZECJP9nZoMGYoE5T": {
    "H": [ {
        "kind": "CLASS_ALIAS",
        "type": {
          "iri": "http://lumii.lv/ontologies/2016/mini-bkus-en#HospitalEpisode",
          "prefix": "",
          "local_name": "HospitalEpisode",
          "display_name": "HospitalEpisode",
          "is_local": true},
        "context": "QZECJP9nZoMGYoE5T"
      }],
    "T_Count": [{
        "kind": "AGGREGATE_ALIAS",
        "type": null,
        "context": "e6xbGNrCirZMiZjtu",
        "upBySubQuery": 1,
        "distanceFromClass": 1
      }],
    "caseRecordNo": [ {
        "kind": "PROPERTY_NAME",
        "type": {
          "iri": "http://lumii.lv/ontologies/2016/mini-bkus-en#caseRecordNo",
          "prefix": "",
          "display_name": "caseRecordNo",
          "local_name": "caseRecordNo",
          "object_cnt": 0,
          "data_cnt": 20137,
          "is_local": true,
          "max_cardinality": 1,
          "inverse_max_cardinality": -1,
          "data_type": "xsd:integer",
          "property_type": "DATA_PROPERTY"},
        "context": "QZECJP9nZoMGYoE5T"
      }]}
```

**Figure 16.** A Symbol Table Fragment example

## 3.3. Expression Structure and Parsing

Essentially, an expression is built by operations from constants (literals or IRIs), node and field references and path expressions that can be property paths (in the most common case – a single property from the data schema) or explicit query variables. To specify the expressions in textual form, a concrete expression syntax is needed. Since the expressions to be specified need to offer the constructs that are to be mapped into the SPARQL language (WEB, b), the SPARQL 1.1 syntax (its part, starting from the "Expression" rule) is used as the basis of the *ViziQuer* expression syntax.

The *ViziQuer* expression grammar is provided in Appendix 1. If compared to the original SPARQL expression grammar, it has the following modifications:

- There are no explicit SPARQL variable terms, built using the '?' notation, in the *ViziQuer* expressions (except for standalone property and class variables not included in richer expression structures).

  In the places where the SPARQL grammar admits a variable, the *ViziQuer* grammar permits using:
  - a reference to a node, a field, or an explicit query variable[7] (to be translated into the corresponding SPARQL variable); such references can optionally include the @ prefix, or
  - a path expression (e.g., a property name), to be translated into:
    - a SPARQL triple relating the expression (initial) reference item via the path expression to a SPARQL variable, and
    - an expression using the created SPARQL variable.
- The class and property names, as well as individual resources can be specified using their unique display name form (syntactically, a text in *[* and *]* brackets, possibly prefixed by a namespace prefix), cf. Section 3.4.
- If a URI in an expression has to be interpreted as a constant URI and not as a property URI denoting its value, ` needs to be used before the URI (as in *`rdf:type*).
- A restriction on arithmetics: the operators +, -, * and / in the expressions need to be surrounded with spaces to distinguish their arithmetic operation meaning from their symbol usage in URIs or property path constructions.
- There are custom shorthand notations introduced to support easier expression creation in different envisioned use cases (cf. also Section 2.6):
  - The property names can be suffixed by a language tag or their group, as in *abc@en* or *abc@(en,de)*, for specifying property value selection in the designated languages only.
  - The path expressions can be specified using the point notation (*H.admissionTime*, *patient.outpatientEpisode*) following a UML-style convention for navigation expressions (the SPARQL-like syntax of using / as the path item separator is allowed, as well).
  - A REGEX expression can be written in an infix form, using SQL-style ~, and ~* (for case insensitive matching) or LIKE operators (so, *REGEX(?ward, "1-2$")*, *ward ~ '1-2$'* and *ward LIKE '%1-2'* all denote equivalent expressions).

---

[7] Note that even in the case of an explicit query variable (that has been introduced using the ? notation), the references to it from the expressions elsewhere are made without the ? prefix.

- BETWEEN notation: the expression *lengthInDays BETWEEN (10, 30)* is a shorthand for *lengthInDays>= 10 && lengthInDays<= 30*.
- The initial substring, e.g., *SUBSTRING(ward,1,3)*, can be specified by its length in the brackets (e.g., *ward[3]*).
- The inverse property specification *^outpatientEpisode* can be written also as *INV(outpatientEpisode)*.
- There is a shorthand notation for data differences that can be used for queries over OpenLink Virtuoso backed data endpoints: *days(x-y)* expresses the same, as *bif:datediff("day", x, y)*; there are functions *years* and *months*, as well.

The provided grammar is used to build a JSON-encoded parse tree for a given expression. The parsing in *ViziQuer* is done by PEG.js library (WEB, a). There are rules added to the grammar terms that identify the name entities within the expression text (in accordance with the places where the SPARQL grammar would have allowed a SPARQL variable, as described above), the parsing shall place each such entity under the *var* key in the created JSON representation (also covering the mark for a node/field reference (@), property as value mark (`) or property path modifier (?, * or +), if there is one coming with the name). These name entities are then further subjected to name resolution and kind/type enrichment, as described in Section 3.4, before storing the entire expression parse tree in the *parse_exp* attribute in the AST.

## 3.4. Name Resolution and Expression Enrichment

To enable the interpretation of expressions and their use in SPARQL query generation, the names used therein must be resolved as references to their definition in the query text, or as direct references to the entities (properties, classes) of the data model (data schema).

The names to be resolved can be of the ***following forms***:

- *@name* – a plain string, prefixed by @, denoting a reference to a name, defined in the query,
- *prefix:name* – a prefixed name, denoting a reference to an element (a class or a property) of the data model, or an unknown URI, if the reference cannot be found, and the prefix is known in the data model,
- *<iri>* - a full iri, put in braces, denoting an element of the data model, or an unknown URI,
- *name* – a plain string that can be interpreted either as a reference to a name or as a reference to a data model element that belongs to the default namespace,
- *[text]* or *prefix:[text]* – text in square brackets, with or without the prefix part, denoting a designated display name (stored in the data model) for a data model entity,
- *`prefix:name* – a prefixed name (with a known *prefix*), prefixed by `, denoting a property as value and not to be referenced as an element (a class or a property) of the data model.

The ***resolution of a name in an expression within AST*** involves setting the *kind* and *type* attributes for the name in the expression. If the name to be resolved is a ***plain string***, then we distinguish two cases[8]:

- if the name coincides with the ***whole expression***, then it is checked, if it matches an available property from the default namespace within the data model; if so, the name is resolved as the found property name; otherwise, it is looked up in the symbol table, if the name is available in the context of this node (avoiding circular references of the name to itself, of course), and
- if the name is a ***proper part of the expression***, it is looked up first in the symbol table; if it is not found there, then it is looked up as a property from the local namespace within the data model.

```
"parsed exp": [
 {"ConditionalOrExpression": [
   {"ConditionalAndExpression": [
     {"RelationalExpression": {
       "NumericExpressionL": {
        "AdditiveExpression": {
         "MultiplicativeExpression": {
          "UnaryExpression": {
           "PrimaryExpression": {
            "var": {
             "name": "caseRecordNo",
             "ref": null,
             "type": {
              "iri":"http://lumii.lv/ontologies/
2016/mini-bkus-en#caseRecordNo",
              "prefix": "",
              "display name": "caseRecordNo",
              "local name": "caseRecordNo",
              "object cnt": 0,
              "data cnt": 20137,
              "is local": true,
              "max cardinality": 1,
              "inverse max cardinality": -1,
              "data type": "xsd:integer",
              "property_type": "DATA_PROPERTY"
             },
             "kind": "DIRECT PROPERTY",
             "PathMod": null,
             "PropertyAsValueMark": null
            },
            "Substring": "",
            "FunctionBETWEEN": null,
            "FunctionLike": null
           }
          },
         "UnaryExpressionList": []
        },
        "MultiplicativeExpressionList": []
}}}},]},]},]
```

a

```
"parsed_exp": [
 {"ConditionalOrExpression": [
   {"ConditionalAndExpression": [
     {"RelationalExpression": {
       "NumericExpressionL": {
        "AdditiveExpression": {
         "MultiplicativeExpression": {
          "UnaryExpression": {
           "PrimaryExpression": {
            "var": {
             "name": "T_Count",
             "ref": null,
             "type": null,
             "kind": "AGGREGATE_ALIAS",
             "PathMod": null,
             "PropertyAsValueMark": null
            },
            "Substring": "",
            "FunctionBETWEEN": null,
            "FunctionLike": null
           }
          },
          "UnaryExpressionList": []
         },
         "MultiplicativeExpressionList": []
}}}},]},]},]
```

b

**Figure 17.** *parse_exp* content examples: *caseRecordNo* (a) and *T_Count* (b) from Figure 16.

For a name resolved as a ***property from the data model*** (in the described cases of the plain string format, and the prefixed name, iri and bracketed text formats), its *kind* is set in the AST as DIRECT_PROPERTY (a kind distinct from any symbol table kinds) and the supplementary context information from the data model is retrieved and placed in its *type* attribute (cf. an example in Figure 17, a).

[8] To avoid ambiguity, @ prefix can be used for the references to query names and *prefix:name* notation for references to the model properties; the plain string option has been offered for convenience. There are further restrictions not allowing the plain string option, if it is viewed as causing too much ambiguity (e.g., when the name matches both an alias defined in the query and a property in the default namespace).

For a name resolved as a ***name defined in the query***, its *kind* and *type* are looked up in the symbol table for the node, filling it beforehand, if necessary.

The records for the name in the symbol table (at different nodes) are filled with the *kind* and *type* at the point of processing the definition of the name within AST (note that a definition of the name occurs through an alias or through a reference to a property that constitutes the entire field expression). At this point the properties of the name in ST including *kind* (PROPERTY_NAME, PROPERTY_ALIAS, BIND_ALIAS or REFERENCE_TO_ALIAS), *model type* (for simple expressions) and *max cardinality* (possibly also for more complex expressions)) are retrieved from its defining expression, and they are also propagated to all places in ST that refer to the name definition.

If a name in an expression ***cannot be resolved*** either as coming from the data model or as a name defined in the query (e.g., when a name in *prefix:name* form does not match the names known in the data model), its *kind* and type in AST are both set to *null*, however, it still can be used in the generated SPARQL query. If such an unresolved name is to be reflected in the symbol table, its *kind* attribute is still set to PROPERTY_ALIAS or PROPERTY_NAME, depending on whether the field with the unresolved expression has an alias or not (its *type* information is *null* also in the symbol table).

Figure 17, b has a simple example of a parse tree involving a name whose *kind* is AGGREGATE_ALIAS.

### 3.5. Query Environment Parameters

The query diagrams (the projects consisting of the diagrams) in ViziQuer are provided with a set of parameters whose values can be set by the user and that can influence the way SPARQL queries are generated from the visual queries (different settings of parameters can lead to different query SPARQL queries and different results). It is possible to customize SPARQL generation in *ViziQuer* through the following parameters:

**DSS schema** – the name of the schema that holds the data model and will be used for name resolution and for querying (used for accessing the appropriate data model).

**Use String Literal Conversion** (the default value is “SIMPLE”):

- SIMPLE – if in a comparison expression one side is a data property with type *xsd:string* and the other side is a string expression, the data property is embedded in *STR* function (*gender="male"* is considered to be *str(gender)="male"*).
- TYPED – if in a comparison expression one side is a data property with type *xsd:string* and the other side is a string expression, the suffix *^^xsd:string* is attached to the string expression (*gender="male"* is considered to be *gender="male"^^xsd:string*).
- OFF – no transformations are done (this may result in overlooking literal equalities SPARQL query execution, as *"male"* and *"male"^^xsd:string* are different values).

**Query Engine Type** – the type GENERAL invokes no expression transformation, while the value VIRTUOSO implies:

- The shorthand expression *days(a-b)* is enabled for date value differences to denote *bif:datediff("day",a,b)*, similar functions *months* and *years* are available, as well; the expressions *a* and *b* are surrounded by *xsd:dateTime(..)*, if they are of the data type *xsd:date*.
- The substrings are implemented using *bif:substring* function.

**Use Default Grouping Separator** – specify the separator used in the aggregation function GROUP_CONCAT. The default value is “, ”.

**Direct Class Membership Role** –direct class membership role used as a predicate value in the instance class triple (shall use *rdf:type*, if not specified).

**Indirect Class Membership Role** – indirect class membership role used as a predicate value in the instance class triple when it is specified in the class to use the indirect role.

**Enable Wikibase Label Services** – specifies whether to enable Wikibase Label Services for selecting labels and descriptions along with attributes in queries over *Wikidata*.

# 4. SPARQL Query Generation

In the next step, the ViziQuer graphical query abstract syntax is translated into a dedicated model for generating SPARQL queries. This translation also considers the symbol table and the parameters that are set by the user. The process starts with SPARQL variable name assignment to the names created and used in the query AST.

## 4.1. SPARQL Variable Names

The principal rule for the SPARQL variable name creation in the case of an explicitly provided name (alias) in the visual query is to prefix the explicit name by ? (so, the name *X* becomes the SPARQL variable *?X*). This implies that if different items in the query are marked by the ***same explicit variable***, they shall correspond also to the ***same SPARQL variable***. In the case if no explicit name is provided for a query item, the name is ***auto-generated*** from the context information, if available (class name for node instances, property name for data model references within the fields (using the last property in the case of a property path)). The auto-generated names, however, are ***kept distinct*** (except in the case of alternate *UNION* branches) by the means of adding appropriate suffixes (so, two uses of a property *:abc* would result in variables *?abc* and *?abc_1*).

Regarding the use of a name in an expression, the name kind in AST distinguishes among the reference to an alias and a used property / data model element (cf. Section 3.4). If the name denotes a data model element (or an unknown property), it introduces a new (locally used) auto-generated name. If the name is an alias, the symbol table is used to find the corresponding SPARQL variable name: each name description in ST at the node where the name is used shall have a corresponding introduction context information (in the case of multiple available contexts, if their SPARQL variable names coincide (this would correspond e.g., to the case of an explicit name introduced in multiple places), take this name; otherwise choose any of names and flag an error).

## 4.2. SPARQL Query Generation Model

The structure of the SPARQL query generation model is outlined in Figure 18. The model is structured along the node, link, and field structure of the visual query, however, the data filled in are collected for the needs of the SPARQL query generation.

The central class ***Node*** – represents the node of the ViziQuer graphical query, with its child node structure and most of the node attributes inherited from the query AST (Figure 13). The node instance variable name (*classInstance* attribute) is generated, as explained in Section 4.1. The *classDefinition* attribute shall contain a textual SPARQL triple, asserting the class name of the node instance. Furthermore, the *linkDefinition* attribute is generated to contain the textual SPARQL form of the link triple connecting the node to its parent node.

The blocks subordinated to the *Node* class contain further information relevant to the SPARQL query generation.

***Select*** contains the variables coming from the node (for nodes not inside negation blocks) that go into the select clause of the query (or the subquery involving the node fragment). The list consists of simple variables, label variables and aggregated variables.

- ***Simple variables*** are created for distinct items in the following list:
  - variables from node attribute fields that are not marked in AST as helpers,
  - target node instances of reference links starting at the node within a subquery and going out of the subquery scope, and
  - explicit references to nodes from outside the subquery scope, made from attribute definitions (both triple and expression forms) and filters at the node.

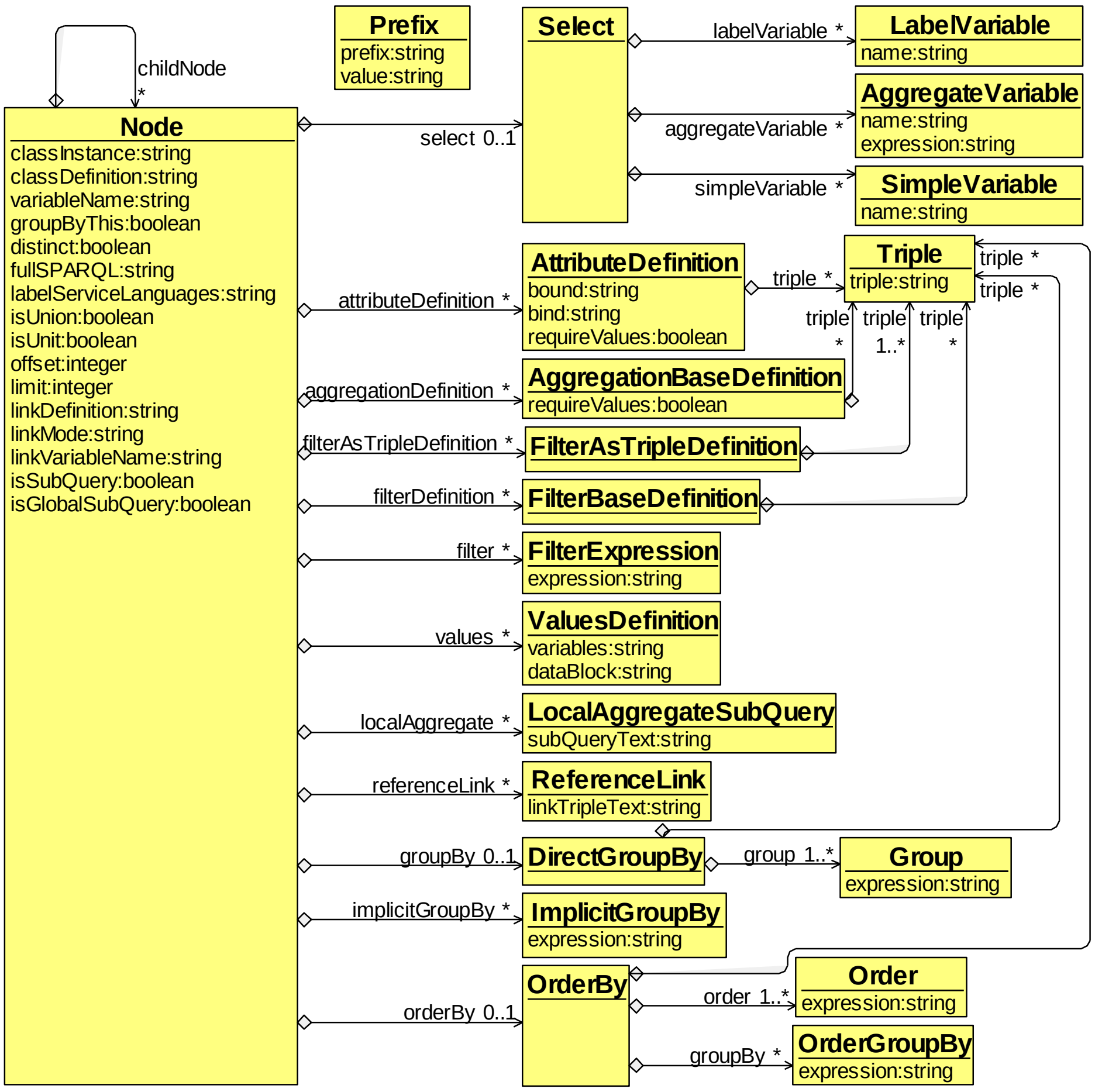

**Figure 18.** SPARQL query generation model

- ***Aggregate variable*** – a variable from a node aggregation field if the node is the main node of the query or a subquery. Each variable has a name and an expression.
- ***Label variables*** are created for attributes marked by *add Label*, *add AltLabel* or *add Description* if the parameter *Enable Wikibase Label Services* is set**.** The label variable name has *Label/AltLabel/Description* suffix added to the base variable name.

***AttributeDefinition*** contains the SPARQL clauses describing simple non-aggregated attribute fields of the node (the instance fields whose expressions involve aggregate functions are reflected in *LocalAggregateSubquery* class instead). It consists of:

- ***triple patterns*** for all properties included in the field expression, connecting their values to the node instance variable or another explicit navigation starting point,
- ***bind clauses*** (formed if the attribute field contains an expression that is more than a property name/property path),
- ***bound clause*** (formed if the field has required values, is implemented by a *bind* clause, and contains an expression not known to necessarily produce a value),
- the ***requireValues*** mark for required attribute values (the triple patterns for an attribute without the mark are to be included in an *OPTIONAL* block).

For instance, at a node *n*, the field with the expression *totalCost - 3* and alias *t* creates a triple pattern *?n :totalCost ?totalCost* and a bind clause *BIND(?totalCost-3 AS ?t)*.

***AggregationBaseDefinition*** contains triple patterns for evaluating the properties within the aggregated function argument expression at an aggregation field of the node. For instance, the expression *avg(totalCost / lengthInDays)* at a node *H* shall lead to triples *?H :totalCost ?totalCost* and *?H :lengthInDays ?lengthInDays*. (Note: computing the expression itself and applying the aggregate function is recorded at *Select* class).

***FilterAsTripleDefinition*** is created when a condition can be expressed in the form of a triple connecting the node instance variable (or another reference point) by a property or a property path to a value (a literal or URI), or a fully introduced (non-optional) name. For instance, if node *H* has a condition *id = 12345*, the triple shall be *?H :id 12345*.

The conditions that do not fit the triple form are expressed using ***FilterBaseDefinition*** and ***FilterExpression*** elements. The ***FilterExpression*** alone is used for conditions that fit into a single *FILTER* clause (as e.g., *FILTER(?T >= 4)* for the before defined variable *?T*) or *FILTER EXISTS* block (as e.g., *FILTER EXISTS{?H :lengthInDays ?lengthInDays. FILTER(?lengthInDays >= 10)}* for expression *lengthInDays >= 10)*.

The ***FilterBaseDefinition*** is used to record the triples to be added to the query outside the *FILTER* expression, as e.g., *?H :lengthInDays ?lengthInDays* outside *FILTER(?lengthInDays >= 10)*; the *FILTER* part is still included in the *FilterExpression*. The decision on whether to use the *FILTER EXISTS* or a simple *FILTER* form (with additional triples) depends on the multiplicity of the property (*:lengthInDays* in the example); either of the options can be enforced manually, as well.

***LocalAggregateSubQuery*** deals with the situation when there is an aggregate expression specified in an ordinary attribute field, invoking an aggregate subquery computation in the context of the node instance. The *subQueryText* attribute of the instance shall contain the entire subquery text, including the *select*, *where* and *group by* clauses. For instance, the field expression *count(id)* at a node *H* shall lead to the subquery text *{SELECT ?H (COUNT(?id) AS ?id_COUNT) WHERE{?H :id ?id.} GROUP BY ?H}*.

***ReferenceLink*** contains the triple pattern reflecting a positive reference link from the current node to a referred node, cf. e.g., Figure 7, the link *hospitalEpisode* between *HospitalEpisode* and *CPhysician* classes.

***DirectGroupBy*** contains grouping expressions. It consists of:

- variable names to be grouped by (the *Group* class), and
- triple patterns for the property names in the group by expression (for instance, *{?H :id ?id}*, for *id* a model property (not a field alias) in the *group by* field of node *H*).

The *GROUP BY* section of the generated SPARQL text (cf. Section 4.4) for aggregated queries, in addition to explicit grouping fields, shall contain references to all non-aggregated elements of the *SELECT* list (gathered in ***ImplicitGroupBy*** class).

***OrderBy*** describes the ordering expression. It consists of:

- *ORDER BY* clause consisting of variables to be ordered, each with the *DESC* label, if specified,
- triple patterns for the properties included in the order by an expression that does not match the fields defined in the query, and
- names of variables introduced in the *order by* expression; to be included also in the *GROUP BY* clause (needs to contain also the non-aggregated ordering variables).

## 4.3. SPARQL Query Model Creation Example

To build the SPARQL generation model tree from the ViziQuer abstract query syntax tree from the AST example in Figure 14, the following steps are taken:

Starting from the model tree root node:

1. A node instance name *?H* is generated from the root node instance name *H*.
2. A triple pattern *?H a :HospitalEpisode.* describing the class assertion of the node is generated from the root node identification information.
3. The query simple select clause variables and the triple patterns describing properties are formed from attribute fields:
    a. The attribute field *caseRecordNo* is a data property from the schema, so a simple variable *?caseRecordNo* is added to the *Select* section of the model into the *SimpleVariable* section, and the attribute definition triple pattern *?H :caseRecordNo ?caseRecordNo* is added to the *AttributeDefinition* section.
    b. The attribute field *T_Count* is a reference to a subquery selection field, so, only a simple variable ?*T_Count* is added to the *Select* block of the model into the *SimpleVariable* section.
4. A subquery clause is generated:
    a. A node instance name *?TreatmentInWard* is generated from the node identification information (the class name, since no alias has been specified).
    b. A triple pattern *?TreatmentInWard a :TreatmentInWard.* for the class assertion of the node is generated from the node identification information.
    c. An incoming link *linkMode = REQUIRED* and *isSubQuery = TRUE* properties are set.
    d. A triple pattern *?H :treatmentInWard ?TreatmentInWard.* describing the link definition is generated from the current and the parent nodes' information.
    e. The query aggregated select clause variables are formed from aggregation fields. (the notation *count(.)* is interpreted as the *COUNT(TreatmentInWard)* expression, so an aggregated variable *COUNT(?TreatmentInWard)* is added to the *Select* section of the model into *AggregateVariable* section. Since the triple pattern describing the class assertion of the node was already defined, no triple patterns are added to the model).

5. A condition *T_Count >= 4* is transformed into a filter expression *FILTER(?T_Count >= 4)* and is added to the *FilterExpression* section of the model. Since variable *T_Count* is a reference from the subquery, no triple pattern is added to the model.
6. The Order by expression *DESC(?T_Count)* is added to the *OrderBy* section.
Figure 19 outlines the described creation of the SPARQL query generation model.

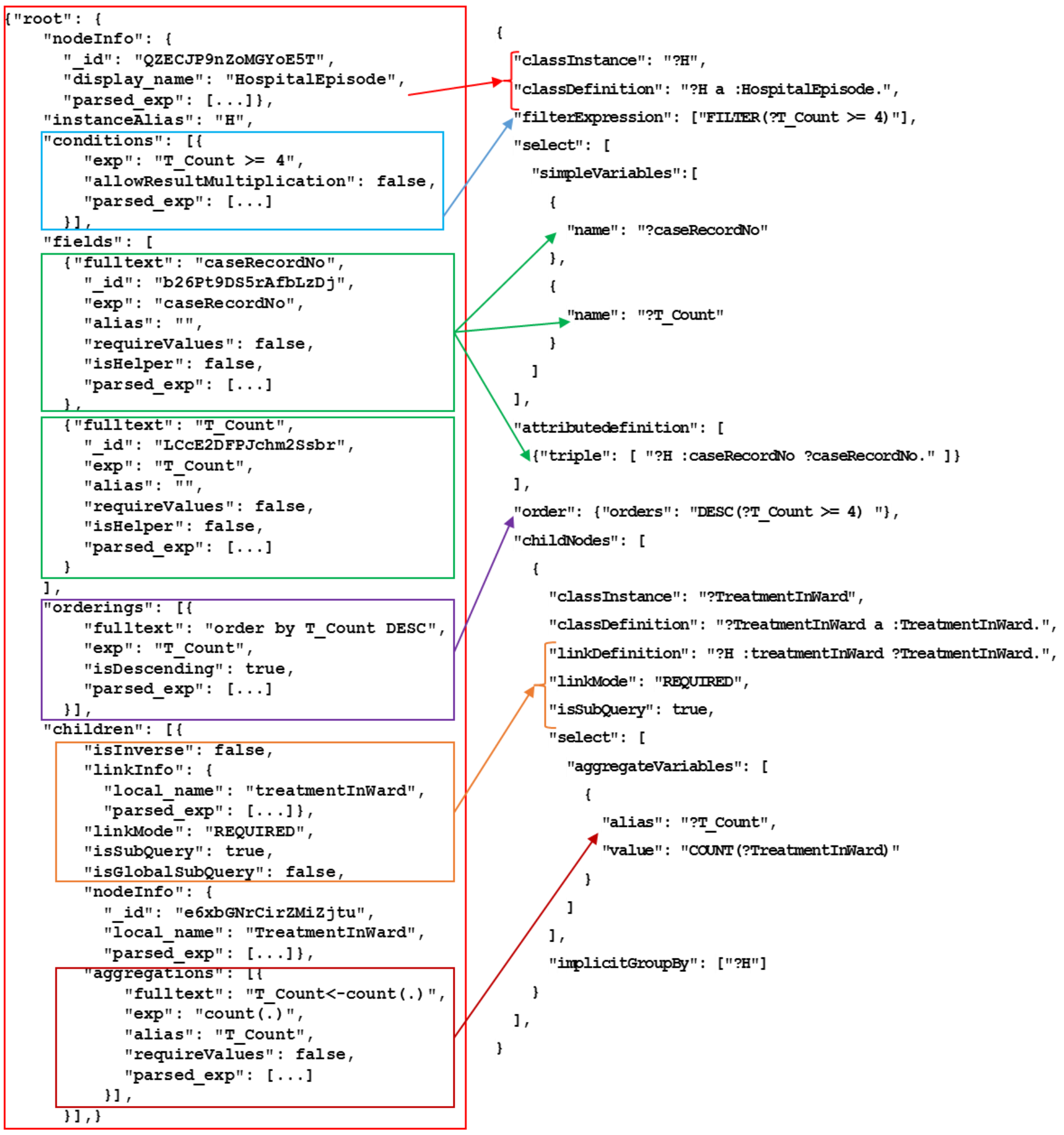

**Figure 19.** ViziQuer query abstract syntax tree and SPARQL query generation model tree

## 4.4. SPARQL Query Text Generation

The last thing to be done is to generate the SPARQL textual query from the SPARQL Query Generation Model.

The SPARQL text is generated in a certain order. First, the definition of all prefixes used in expressions in the query are retrieved from the data model and are placed in the SPARQL text. The SPARQL query SELECT clause is then generated to include:

- Simple variables from all nodes that are not under negation or subquery links.

- Aggregated variables from all nodes that are not under negation or subquery links.
- Label variables from all nodes that are not under negation or subquery links.

The query WHERE clause is built by collecting its textual form via traversing the SPARQL query abstract syntax tree, starting from the root node, collecting for each query fragment (set of nodes connected by required join links) the SPARQL code in the following order, split over the three phases:

- **Phase 1.** Initial Structure:
    - Incoming links into fragment nodes, node class assertions and positive reference links and filters as triples among the fragment nodes and to nodes above the fragment.
    - The first required attribute of any data node in the fragment, not introduced so far (to have the node variable defined).
    - Grounding: re-introduction of names defined in the query above the fragment, used in the fragment and not defined in the fragment.

- **Phase 2.** Required and optional subqueries attached to the fragment nodes, as well as Direct SPARQL texts of the subquery form, ascribed to the fragment nodes.

- **Phase 3.** Main Traversal: For each fragment node, starting from the fragment head node, in a depth-first search manner:
    - Processing of child nodes (in-fragment nodes) with required join links,
    - Processing of UNION group fragments behind the node,
    - Processing of fragments behind optional join links from the node,
    - Attribute definitions (including triples and BIND/BOUND expressions) and VALUES clauses, in the order of their visual appearance,
    - Full SPARQL texts without subquery, attached to the node (the introduced names can be used e.g., in filters),
    - Filters as triple expressions (the ones not processed in Phase 1),
    - Filter base triple definitions,
    - At the fragment head node: filter expressions (including *filter exists* and *filter not exists* expressions) for all nodes from the fragment,
    - Negated subqueries (MINUS queries) attached to the node,
    - Aggregation base triple definitions (for fragment head node only).

At the end of processing the main query or a subquery, the triple pattern definitions from the ORDER BY and the GROUP BY expressions are added to the SPARQL text and the GROUP BY, ORDER BY, OFFSET and LIMIT clauses themselves are added.

We note that the order of collecting the SPARQL code fragments is important for the correct placement of place sensitive SPARQL constructs (e.g., MINUS or BIND), as well as a reasonable order of patterns in the query may enhance the readability of the created SPARQL queries and help the query execution engines to build a reasonable plan for query execution. The proposed and implemented SPARQL query generation order creates a particular SPARQL order; the possibilities of the visual query notation to encode a different pattern order in the SPARQL query may be limited. Ways of relaxing these limitations are a subject of future work and are beyond the scope of this paper.

Figure 20 shows an example of the textual SPARQL query generated from the used SPARQL abstract syntax structure of the query of Figure 19.

```
PREFIX : <http://lumii.lv/ontologies/2016/mini-bkus-en#>
SELECT ?caseRecordNo ?T_count WHERE{
  ?H a :HospitalEpisode.
  {SELECT ?H (COUNT(?TreatmentInWard) AS ?T_count) WHERE{
    ?H :treatmentInWard ?TreatmentInWard.
    ?TreatmentInWard a :TreatmentInWard.}
    GROUP BY ?H }
  OPTIONAL{?H :caseRecordNo ?caseRecordNo.}
  FILTER(?T_count >= 4)
} ORDER BY DESC(?T_count)
```

**Figure 20.** Textual SPARQL query corresponding to Figure 19 abstract structure.

## 5. Conclusions

We have demonstrated the feasibility of implementation of a rich visual query language over RDF data via its translation into the standard textual SPARQL query language. The source of the translation is an encoding of visual diagrams in the form of boxes, lines, and compartments, as provided by the generic visual DSL platform *ajoo* (Sprogis, 2016) and the translation target is the rich textual structure of SPARQL. The principal milestones of the translation involve creation of an abstract query model, followed by a dedicated model for generating the textual SPARQL code out of the abstract query model structure.

The identified query translation structures and steps separate the translation into self-standing blocks that can allow for being reused within other tasks of serving the visual query diagrams, as, e.g., using the abstract syntactic structure and the symbol table for context-based query completion, or reversely generating the visual presentation of an existing SPARQL query (the prototype functionality corresponding to both these use cases is available in the existing *ViziQuer* visual tool environment). There would also be a re-use possibility in creating other implementations of the visual query language (for instance, by translating the visual queries directly into SQL for execution over data back-ends that are supported by relational databases).

The described implementation solution can support a rich set of visual queries that involve optional and negated links, as well as aggregation and subqueries, and advanced expressions for the data selection and conditions. The visual notation permits reverse visualization of a rich set of SPARQL queries, as well. To be able to fine-tune the reverse translation SPARQL queries back into the visual notation, the visual notation still needs to be further enriched to include means for query clause ordering specification; we expect that a natural extension of the provided notation would be able to cope with the task. Another work in progress item relates to the named graph specification options that would require the expansion of the provided notation.

## Acknowledgements

This work has been carried through at Institute of Mathematics and Computer Science, University of Latvia, and has been partially supported by Latvian Science Council grant lzp-2021/1-0389 “Visual Queries in Distributed Knowledge Graphs” and project “Strengthening of the capacity of doctoral studies at the University of Latvia within the framework of the new doctoral model”, identification No. 8.2.2.0/20/I/006 (J.Ovčiņņikova).

# Appendix 1 – ViziQuer Expression Grammar[9]

```
Expression = "[ ]" / "[ + ]" / "(no_class)" / ValueScope / ConditionalOrExpression /
ClassExpr /"*"
ValueScope = "{" (ValueScopeA / ValueScopeB / ValueScopeC) "}"
ValueScopeA = INTEGER ".." INTEGER
ValueScopeB = Scope ("," Scope)*
ValueScopeC = (UNDEF / PrimaryExpression) ("," (UNDEF / PrimaryExpression))*
Scope = "(" ((UNDEF / PrimaryExpression) ("," (UNDEF / PrimaryExpression))*) ")"
ClassExpr = "(.)" / "." / "(select this)" / "(this)"
ConditionalOrExpression = ConditionalAndExpression (("||"/ "OR") ConditionalAndExpression )*
ConditionalAndExpression = RelationalExpression (("&&" / "AND") RelationalExpression )*
RelationalExpression = RelationalExpressionA / RelationalExpressionB / RelationalExpressionC
         / RelationalExpressionD
RelationalExpressionA = NumericExpression ("IN" / "NOT" "IN") (ExpressionListA /
ExpressionListB / ExpressionListC)
RelationalExpressionB = ClassExpr Relation NumericExpression
RelationalExpressionC = NumericExpression Relation ClassExpr)
RelationalExpressionD = NumericExpression (Relation NumericExpression)?
NumericExpression = AdditiveExpression
AdditiveExpression = MultiplicativeExpression MultiplicativeExpressionList*
MultiplicativeExpressionList = Concat / Additive / NumericLiteralPositive /
         NumericLiteralNegative
Concat = "++"  MultiplicativeExpression
Additive = ("+" / "-") MultiplicativeExpression
MultiplicativeExpression = UnaryExpression UnaryExpressionList*
UnaryExpression = ("!" / "-" )? PrimaryExpression
UnaryExpressionList = ("*" / "/") UnaryExpression
PrimaryExpression = BooleanLiteral / BuiltInCall / QName / iriOrFunction / RDFLiteral /
         BrackettedExpression / NumericLiteral / Var / DoubleSquareBracketName / LN
PrimaryExpressionCOALESCE = BooleanLiteral / iriOrFunction / BuiltInCallNoFunction /
         RDFLiteral / BrackettedExpression / NumericLiteral / Var / DoubleSquareBracketName
         / QName / LN
BooleanLiteral = "true" / "false"
RDFLiteral = StringQuotes (LANGTAG / ("^^" iri))?
BrackettedExpression = "(" Expression ")"
BuiltInCall = Aggregate / FunctionExpression / RegexExpression / SubstringExpression /
         SubstringBifExpression / StrReplaceExpression / ExistsFunc / NotExistsFunc
BuiltInCallNoFunction = Aggregate / RegexExpression / SubstringExpression /
         SubstringBifExpression / StrReplaceExpression / ExistsFunc / NotExistsFunc
Aggregate = AggregateA / AggregateB / AggregateC
AggregateA = "COUNT_DISTINCT" "(" Expression ")"
AggregateB = ("COUNT"/ "SUM"/ "MIN"/ "MAX"/ "AVG"/ "SAMPLE") "(" "DISTINCT"? Expression ")"
AggregateC = "GROUP_CONCAT" "(" "DISTINCT"? Expression SEPARATOR? ")"
SEPARATOR = (";" "SEPARATOR" "=" StringQuotes ) / ("," (StringQuotes))
FunctionExpression = FunctionExpressionC / FunctionExpressionA / FunctionExpressionB /
         IFFunction / FunctionExpressionD / FunctionExpressionLANGMATCHES / FunctionCOALESCE
         / BOUNDFunction / NilFunction / BNODEFunction
FunctionExpressionA = ("LANG" / "DATATYPE" / "IRI" / "URI" / "ABS" / "CEIL" / "FLOOR" /
         "ROUND" / "STRLEN" / "STR" / "UCASE" / "LCASE" / "ENCODE_FOR_URI" / "YEAR" /
         "MONTH" / "DAY" / "HOURS" / "MINUTES" / "SECONDS" / "TIMEZONE" / "TZ" / "MD5" /
         "SHA1" / "SHA256" / "SHA384" / "SHA512" / "isIRI" / "isURI" / "isBLANK" /
         "dateTime" / "date" / "isLITERAL" / "isNUMERIC") "(" Expression ")"
FunctionExpressionB = ("LANGMATCHES" / "CONTAINS" / "STRSTARTS" / "STRENDS" / "STRBEFORE" /
         "STRAFTER" / "STRLANG" / "STRDT" / "sameTerm") "(" Expression "," Expression ")"
FunctionExpressionC = ("days" / "years" / "months" / "hours" / "minutes" / "seconds" ) "("
         PrimaryExpression "-" PrimaryExpression ")"
FunctionExpressionD = ("COALESCE" / "CONCAT") ExpressionListA
```

[9] The actual implementation uses a wider range of characters for literals and variable names, allowing Unicode in delimited string constants, as well as variable names, surrounded by [ and ].

```
FunctionCOALESCE = PrimaryExpressionCOALESCE "??" PrimaryExpressionCOALESCE
FunctionExpressionLANGMATCHES = (PNAME_LN / QName / LN) (LANGTAG_MUL / LANGTAG)
BOUNDFunction = "BOUND" "(" PrimaryExpression")"
NilFunction = ("RAND" / "NOW" / "UUID" / "STRUUID") NIL
BNODEFunction = "BNODE" (("(" Expression ")") / NIL)
IFFunction = "IF" "(" Expression "," Expression "," Expression ")"
RegexExpression = "REGEX" "(" Expression  "," Expression ("," Expression)? ")"
SubstringExpression = ("SUBSTRING"/"SUBSTR" ) "(" Expression "," Expression (","
         Expression)? ")"
SubstringBifExpression = ("bif:SUBSTRING" / "bif:SUBSTR" ) "(" Expression  "," Expression
         ("," Expression)? ")"
StrReplaceExpression = "REPLACE" "(" Expression  "," Expression ("," Expression)? ")"
ExistsFunc = ExistsFuncA / ExistsFuncB
ExistsFuncA = "EXISTS" ("(" Expression ")") / Expression
ExistsFuncB = "{" Expression "}"
NotExistsFunc = NotExistsFuncA / NotExistsFuncB
NotExistsFuncA = "NOT" "{" Expression "}"
NotExistsFuncB = "NOT" "EXISTS"? ("(" Expression ")") / Expression
ExpressionListA = NIL / "(" Expression  ( "," Expression )* ")"
ExpressionListB = "{" Expression  ( "," Expression )* "}"
ExpressionListC = "{" INTEGER ".." INTEGER "}"
LANGTAG = "@" StringLang
LANGTAG_MUL = "@" "(" (StringLang ("," StringLang)*) ")"
iri = IRIREF / PNAME_LN
IRIREF = "<" ([A-Za-z_0-9:.#/-()%,] / "\\")* ">"
PNAME_NS = (PN_PREFIX? ":")
PNAME_LN = "@"? "`"? PNAME_NS (CharsStringVarname / BasicCharStringNum)
         SubstringBetweenLikeExpression
PN_PREFIX = BasicCharString
iriOrFunction = iri ArgList?
ArgList = ("(" "DISTINCT"? ArgListExpression ")" ) / NIL
NIL = "(" ")"
ArgListExpression = Expression ( "," Expression )*
NumericLiteral = NumericLiteralUnsigned / NumericLiteralPositive / NumericLiteralNegative
NumericLiteralUnsigned = DOUBLE / DECIMAL / INTEGER
NumericLiteralPositive = DECIMAL_POSITIVE / DOUBLE_POSITIVE / INTEGER_POSITIVE
NumericLiteralNegative = DECIMAL_NEGATIVE / DOUBLE_NEGATIVE / INTEGER_NEGATIVE
DECIMAL = [0-9]* "." [0-9]+
DOUBLE = ([0-9]+"."[0-9]*[eE][+-]?[0-9]+)/("."([0-9])+[eE][+-]?[0-9]+)/(([0-9])+ [eE] [+-]?
         [0-9]+)
INTEGER = [0-9]+
INTEGER_POSITIVE = ("+" INTEGER)
DECIMAL_POSITIVE = ("+" DECIMAL)
DOUBLE_POSITIVE = ("-" DOUBLE)
INTEGER_NEGATIVE = ("-" INTEGER)
DECIMAL_NEGATIVE = ("-" DECIMAL)
DOUBLE_NEGATIVE = ("-" DOUBLE)
Var = ("??" VARNAME?) / ("?" VARNAME) / ("$" VARNAME)
StringQuotes = STRING_LITERAL1 / STRING_LITERAL2
STRING_LITERAL1 = "'" QuotedString "'"
STRING_LITERAL2 = doubleQuotes QuotedString doubleQuotes
QName = Path / PathBrRound /PathBr
Path = PathAlternative SubstringBetweenLikeExpression
PathBrRound = "[[(" PathAlternativeBr SubstringSpec? ")]]" BetweenExpression?
LikeExpression?
PathBr = "[[" PathAlternativeBr SubstringSpec? "]]" BetweenExpression? LikeExpression?
PathAlternative = PathSequence ("|" PathSequence)*
PathAlternativeBr = PathSequenceBr ("|" PathSequenceBr)*
PathSequence = PathEltOrInverse (("." / "/") PathEltOrInverse)+
PathSequenceBr = PathEltOrInverse (("." / "/") PathEltOrInverse)*
PathEltOrInverse = PathEltA / PathEltB
PathEltB = "^"? PathElt
```

```
PathEltA = "inv" "(" PathElt ")"
PathElt = PathPrimary PathMod?
PathPrimary = ("!" PathNegatedPropertySet) / iriP / ("(" Path ")") / LNameP/ "a"
PathNegatedPropertySet = PathNegatedPropertySetBracketted / PathOneInPropertySet
PathNegatedPropertySetBracketted = "("(PathOneInPropertySet("|" PathOneInPropertySet)*)? ")"
PathOneInPropertySet = PathOneInPropertySet2 / PathOneInPropertySet1
PathOneInPropertySet1 = "^"? (iriP / LNameP/ 'a')
PathOneInPropertySet2 = "inv" "(" (iriP / LNameP/ 'a' ) ")"
iriP = IRIREF / PNAME_LNP / PNAME_NSP
PNAME_NSP = PN_PREFIX? ':'
PNAME_NSPSQ = BasicCharString ":"
PNAME_LNP = "@"? PNAME_NSP (BracketedCharString / BasicCharString)
LNameP = "@"? (BracketedCharString / BasicCharString)
LN = LNameINV / LNameINV2 / LName
LNameSimple = CharsStringVarname / BasicCharString
LName = "@"? "`"? LNameSimple PathMod? SubstringBetweenLikeExpression
LNameINV = "@"? "`"? "INV" "(" PNAME_NS? LNameSimple ")" SubstringBetweenLikeExpression
LNameINV2 = "@"? "`"? "^" LNameSimple SubstringBetweenLikeExpression
DoubleSquareBracketName = "`"? PNAME_NSPSQ? BracketedCharString
CharsStringVarname = "[[" BasicCharString "]]"
SubstringBetweenLikeExpression = SubstringSpec? BetweenExpression? LikeExpression?
SubstringSpec = "[" (INTEGER ("," INTEGER)?) "]"
BetweenExpression = "BETWEEN" "(" NumericExpression "," NumericExpression ")"
LikeExpression = LikeExpressionA / LikeExpressionB
LikeExpressionA = ("LIKE" / "~*" / "~") (likeStringA / likeStringB)
LikeExpressionB = ('~*' / '~') (likeStringC / likeStringD)
likeStringA = doubleQuotes "%"? ([A-Za-z0-9_])+ "%"? doubleQuotes
likeStringB = "'" "%"? ([A-Za-z0-9_])+ "%"? "'"
likeStringC = doubleQuotes ([A-Za-z0-9_^* .])+ doubleQuotes
likeStringD = "'" ([A-Za-z0-9_^* .])+ "'"
doubleQuotes = '"'/ '“' / '”'
BasicCharString = [A-Za-z_-] [A-Za-z0-9_-]* ((".") [0-9]*)?
BasicCharStringNum = [A-Za-z0-9_-] [A-Za-z0-9_-]* ((".") [0-9]*)?
BracketedCharString = "[" [A-Za-z0-9_ ] ([A-Za-z0-9_., -()/] / "'")* "]"
StringLang = [A-Za-z]+
QuotedString = ([A-Za-z0-9] / "\\" / "[" / "]" / [-_.:, ^$()!@#%&*+?|/])*
VARNAME = [A-Za-z_] [A-Za-z0-9_]*
Relation = "=" / "!=" / "<>" / "<=" / ">=" /"<" / ">"
PathMod = "?" / "*" / "+"
```